\documentclass[twocolumn,superscriptaddress,letter]{revtex4}%
\usepackage{amssymb}
\usepackage{color}
\usepackage{graphicx}
\usepackage{dcolumn}
\usepackage{bm}
\usepackage[header,title,page,titletoc]{appendix}
\usepackage{amsmath}
\usepackage{amsfonts}%
\providecommand{\U}[1]{\protect \rule{.1in}{.1in}}

\begin{document}
\title{Topological superfluid in a fermionic bilayer optical lattice}
\author{Ya-Jie Wu}
\affiliation{Department of Physics, Beijing Normal University, Beijing 100875, China}
\author{Jing He}
\affiliation{Department of Physics, Beijing Normal University, Beijing 100875, China}
\author{Chun-Li Zang}
\affiliation{Department of Physics, Beijing Normal University, Beijing 100875, China}
\author{Su-Peng Kou}
\thanks{Corresponding author}
\email{spkou@bnu.edu.cn}
\affiliation{Department of Physics, Beijing Normal University, Beijing 100875, China}

\begin{abstract}
In this paper, a topological superfluid phase with Chern number $\mathcal{C}%
=\pm1$, possessing gapless edge states and non-Abelian anyons is designed in a
$\mathcal{C}=\pm1$ topological insulator proximity to an $s$-wave superfluid
on an optical lattice with the effective gauge field and layer-dependent
Zeeman field coupled to ultracold fermionic atoms pseudo spin. We also study
its topological properties and calculate the phase stiffness by using the
random-phase-approximation\ approach. Finally we derive the temperature of the
Kosterlitz-Thouless transition by means of renormalized group theory. Owning
to the existence of non-Abelian anyons, this $\mathcal{C}=\pm1$ topological
superfluid may be a possible candidate for topological quantum computation.

\end{abstract}
\maketitle

\section{Introduction}

Topological quantum computation, based on the manipulation of non-Abelian
anyons\cite{ki1,free}, is considered as an effective method to deal with
decoherence in realizing quantum computation. The first proposed candidate is
the fractional quantum Hall state at filling factor $\nu=5/2$ in ultra
high-mobility samples \cite{eis,xia}. Other proposals are based on two
dimensional (2D) chiral $p_{x}+\mathrm{i}p_{y}$ superconductors with
$\mathcal{C}=\pm1$(the Chern-number) topological invariable\cite{da} and then
the s-wave-superconductor-topological-insulator systems\cite{al,fu}. Among
these approaches, accurate manipulations of single quasi-particles remains a
major difficulty and new techniques are to be expected to overcome this drawback.

On the other hand, cold atoms in optical lattices are an extensively
developing research field\cite{Greiner,jak}. Because one can precisely
controls over the system parameters and defect-free properties, ultracold
atoms in optical lattices provide an ideal platform to study many interesting
physics in condensed matters\cite{Lewenstein,blo}. Some pioneering works
revealed the promising potential of applying ultracold atoms to make quantum
computer and quantum simulator. Recently, experimental realizations of quantum
many-body systems in optical lattices have led to a chance to simulate
strongly correlated electronic systems. By changing the intensity, phase and
polarization of incident laser beams, one can tune the Hamiltonian parameters
including the dimension, the hopping strength and the particle interaction at will.

In this paper, we propose a scenario in which a topological phase, possessing
gapless edge states and non-Abelian anyons, is realized by proximity effect
between a $\mathcal{C}=\pm1$ topological insulator and an $s$-wave superfluid
(SF) of ultracold fermionic atoms in an bilayer optical lattice with an
effective gauge field and a layer-dependent Zeeman field generated by
laser-field\cite{juz,shao,spi,lin,lin1,lin3,blo2}. At the beginning, we give
an effective design of the bilayer Haldane model. Then we put two-component
(two pseudo-spins) interacting fermions on this bilayer optical lattice with
fixed particle concentration. For layer-1, the Haldane model of two-component
fermions at $1/4$ filling under a strong Zeeman field becomes a $\mathcal{C}%
=\pm1$ topological insulator. While for layer-2, there is no Zeeman fields, we
get an s-wave SF state by tuning the interaction between fermion via Feshbach
resonance technique. Thus due to the proximity effect, we get an effective
$\mathcal{C}=\pm1$ topological SF state, of which the $\pi$-flux obeys
non-Aelian statistics and becomes a non-Aelian anyon. Thus, this
$\mathcal{C}=\pm1$ topological superfluid (TSF) may be a possible candidate
for topological quantum computation.

The paper is organized as follows. In Sec. II, we start with the Hamiltonian
of the interacting spinful Haldane model on bilayer optical lattice. In Sec.
III, we calculate the SF order parameter with mean field approach and get a
global phase diagram at zero temperature. In Sec. IV, we point out that there
exists a $\mathcal{C}=\pm1$ TSF\textrm{ }due to the proximity effect of the SF
order in layer-2 on $\mathcal{C}=\pm1$ topological insulator in layer-1. In
Sec. V, we discuss the quantum properties of the $\mathcal{C}=\pm1$ TSF,
including the statistics of the $\pi$-flux and the edge states. In Sec. VI, by
using random-phase-approximation (RPA), we calculate the phase stiffness of
the $\mathcal{C}=\pm1$ topological SF. In Sec.VII, we get the
Kosterlitz-Thouless (KT) transition temperature by the renormalized group (RG)
theory. Finally, the conclusions are given in Sec. VIII.

\begin{figure}[ptb]
\includegraphics[width=0.5\textwidth]{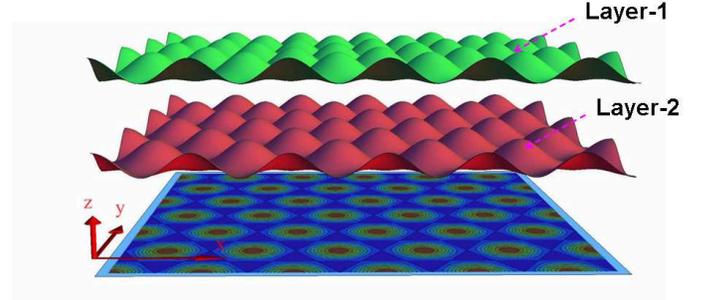}\caption{The illustration of
bilayer honeycomb optical lattice. }%
\end{figure}

\section{The spinful Haldane model on a bilayer optical lattice}

In the first step, we design a bilayer optical lattice of the Haldane model.
In Refs.\cite{shao}, the monolayer optical lattice of the Haldane model had
been proposed in the cold atoms with three blue detuned standing-wave lasers,
of which the optical potential is given by
\begin{equation}
V(x,y)=\sum_{j=1,2,3}V_{0}\sin^{2}[k_{L}(x\cos \theta_{j}+y\sin \theta_{j}%
)+\pi/2] \label{eq1}%
\end{equation}
where $V_{0}$ is the potential amplitude, $\theta_{1}=\pi/3,$ $\theta_{2}%
=2\pi/3,$ $\theta_{3}=0$, and $k_{L}$ is the optical wave vector in XY
plane\cite{zhus}. On the other hand, to design a bilayer optical lattice, the
optical potential from the interference of two opposite-traveling
standing-wave laser beams along the $z$ direction is added as
\begin{equation}
V(z)=V_{L}\sin^{2}(k_{L}^{z}z)-V_{S}\sin^{2}(2k_{L}^{z}z)
\end{equation}
where $V_{L}$ and $V_{S}$ are the amplitudes of the long and short laser along
the z-direction. $k_{L}^{z}$ is the optical wave vector in z-direction. Thus
the total optical potential of the bilayer honeycomb lattice in our case can
be written as
\begin{equation}
V(x,y,z)=V(x,y)+V(z).
\end{equation}
See the illustration in Fig.1. Since the potential barrier of the optical
lattice along the $z$ direction is a double well (See Fig.2), the vertical
tunneling between different bilayer systems is suppressed seriously, each
bilayer can be regarded as an independent 2D honeycomb lattice. The positions
with a minimum potential along z-direction are $k_{L}^{z}z=2\pi n\pm
\arccos \left[  \sqrt{V_{L}/\left(  4V_{s}\right)  }\right]  $ where $n$ is an
integer number.

\begin{figure}[ptbh]
\includegraphics[width=0.3\textwidth]{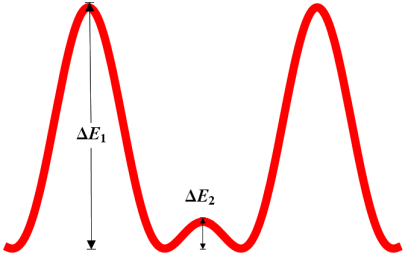}\caption{The optical lattice
potential along z-direction as a double well potential for $\Delta E_{1}%
\gg \Delta E_{2}$.}%
\end{figure}

Let's calculate the tight-binding model of the bilayer optical lattice.
Firstly we derive the{\ hopping parameter} ${t}_{\perp}$ between two layers.
From Fig.2, for $0<V_{L}\leq4V_{S}$ and $\Delta E_{1}=V_{L}+\frac{\left(
4V_{S}-V_{L}\right)  ^{2}}{16V_{S}}\gg \Delta E_{2}=\frac{\left(  4V_{S}%
-V_{L}\right)  ^{2}}{16V_{S}}$ or $\left(  4V_{S}-V_{L}\right)  \rightarrow0$,
one can see that the {optical lattice potential along z-direction can be
}approximately reduced into a double well potential around $z=0$. Then we can
expand $V(z)$ at $z=0$ and get
\begin{equation}
V(z)\simeq \frac{16V_{S}-V_{L}}{3}\left(  k_{L}^{z}z\right)  ^{4}-\left(
4V_{S}-V_{L}\right)  \left(  k_{L}^{z}z\right)  ^{2}.
\end{equation}
We denote $\left \vert 0\right \rangle _{+}$ and $\left \vert 0\right \rangle
_{-}$ as the two nearly degenerate ground states of the double well in the
right-hand and left-hand wells, respectively. The corresponding eigenstates of
the Hamiltonian are odd and even states $\left \vert 0\right \rangle _{e}$ and
$\left \vert 0\right \rangle _{o}$ which are superposition of $\left \vert
0\right \rangle _{\pm}$ such that $\left \vert 0\right \rangle _{o}=1/\sqrt
{2}\left(  \left \vert 0\right \rangle _{+}-\left \vert 0\right \rangle
_{-}\right)  $, and $\left \vert 0\right \rangle _{e}=1/\sqrt{2}\left(
\left \vert 0\right \rangle _{+}+\left \vert 0\right \rangle _{-}\right)  $ with
eigenvalues $E_{0}\pm \Delta E_{0}/2,$ respectively. $\Delta E_{0}$ is the
splitting of the energy levels due to the quantum tunneling effect. We
identify $\Delta E_{0}$, i.e., the hopping parameter ${t}_{\perp}$.

According to the instanton approach\cite{col,jr}, we obtain the instanton
solution as
\begin{equation}
z_{cl}\left(  \tau \right)  \equiv \pm \frac{1}{k_{L}^{z}}\sqrt{\frac{3\left(
4V_{S}-V_{L}\right)  }{2\left(  16V_{S}-V_{L}\right)  }}\tanh \left[
\frac{\omega_{\perp}}{2}\left(  \tau-\tau_{0}\right)  \right]  \label{inst}%
\end{equation}
and then get the energy level splitting $\Delta E_{0}${\ that corresponds to
}${t}_{\perp}$ as\cite{col,jr}%
\begin{equation}
t_{\perp}=\Delta E_{0}/2=2\sqrt{\frac{3(S_{c})_{\perp}}{2\pi}}\omega_{\perp
}e^{-(S_{c})_{\perp}}%
\end{equation}
where the trapping frequency $\omega_{\perp}$ is
\begin{equation}
\omega_{\perp}=\sqrt{\frac{\partial^{2}V(z)}{m\partial z^{2}}\mid_{z=z_{\min}%
}}=\sqrt{8\left(  4V_{S}-V_{L}\right)  E_{r}^{z}},\text{ }%
\end{equation}
with $z_{\min}$ being the value of $z$ when $V(z)$ is a minimal value and the
classical action $(S_{c})_{\perp}$ of instanton is
\begin{equation}
(S_{c})_{\perp}=\frac{\sqrt{2}\left(  4V_{S}-V_{L}\right)  }{\left(
16V_{S}-V_{L}\right)  }\sqrt{\frac{\left(  4V_{S}-V_{L}\right)  }{E_{r}^{z}}}.
\end{equation}
$E_{r}^{z}=\frac{\left(  k_{L}^{z}\right)  ^{2}}{2m}$ is the recoiling energy
of the atoms in the {z-direction} where $m$ is the mass of atoms. We have set
$\hbar=1$.

Secondly we calculate the nearest neighbor {hopping} ${t}$.\ The optical
lattice potential on XY plane is $V(x,y)$ as shown in Eq. (\ref{eq1}) which
forms a honeycomb lattice and simulate the Haldane Model. Substituting
$\theta_{1}=\frac{\pi}{3}$, $\theta_{2}=\frac{2\pi}{3}$, $\theta_{3}=0$ in
$V(x,y)$, we may get%
\begin{equation}
V(x,y)=\frac{V_{0}}{2}\{3+\cos[\kappa(\frac{\sqrt{3}y+x}{2})]+\cos
[\kappa(\frac{\sqrt{3}y-x}{2})]+\cos[\kappa x]\}
\end{equation}
with $\kappa=2k_{L}.$ Around the site A or B in Fig.3, we find $V(x,y)=3V_{0}%
^{2}\kappa^{2}(x^{2}+y^{2})/16-3V_{0}/4$ , i.e.,
\begin{equation}
V(x,y)=m\omega_{0}^{2}r^{2}/2-3V_{0}/4
\end{equation}
with $\omega_{0}=\sqrt{3V_{0}E_{r}^{L}}$, where $E_{r}^{L}=k_{L}^{2}/(2m)$.
Here we may use the the Wentzel, Kramers and Brillouin (WKB) method that has
been extended to two dimensions to estimate $t$ for honeycomb lattice. From
the Ref. \cite{Kean},\ the semiclassical estimation of the tunneling amplitude
reads
\begin{equation}
t=0.707(\frac{V_{0}}{E_{r}^{L}})^{3/4}e^{-0.685\sqrt{\frac{V_{0}}{E_{r}^{L}}}%
}E_{r}^{L}.
\end{equation}

\begin{figure}[ptbh]
\includegraphics[width=0.4\textwidth]{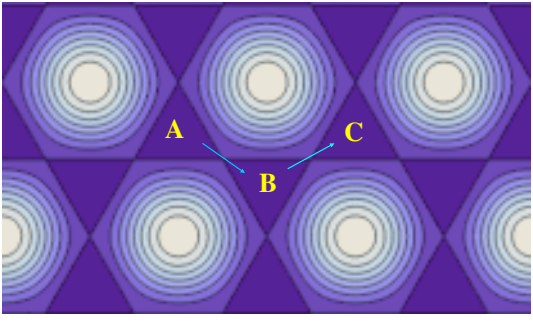}\caption{The instanton process
from A to C that leads to a next nearest neighbor hopping, ${t}^{\prime}$.}%
\end{figure}

Finally we calculate the next nearest neighbor hopping ${t}^{\prime}$. The
situation here is much different from that of $t$ and $t_{\perp}$. The energy
level splitting between two wells (denoted by $A$ and $C$) shown in Fig.3 is
due to the quantum tunneling process from $A$ to $B$ then to $C$. Now we
obtain ${t}^{\prime}$ to be
\begin{equation}
t^{^{\prime}}\simeq \sqrt{2}te^{-S_{c}}=(\frac{V_{0}}{E_{r}^{L}})^{3/4}%
e^{-1.370\sqrt{\frac{V_{0}}{E_{r}^{L}}}}E_{r}^{L}.
\end{equation}
The coefficient $2$ in $e^{-2S_{c}}$ comes from twice of single instanton from
one site to the nearest neighbor site. In this sense, we have
\begin{equation}
t^{^{\prime}}/t\simeq \sqrt{2}e^{-S_{c}}=1.414e^{-0.685\sqrt{V_{0}/E_{r}^{L}}}%
\end{equation}
which is always a small value. For example, if we set $V_{0}=23.806E_{r}^{L},$
the hopping parameters are given to be $t=0.269E_{r}^{L}$ and $t^{\prime
}=0.05t.$

Thus when two-component fermions are put into this bilayer honeycomb optical
lattice, we may get an effective bilayer Haldane model by applying the Raman
laser beams. Taking the tight-binding limit, we can superpose the Bloch states
to obtain eight sets of Wannier functions $w_{A/B,\alpha,\sigma}%
(\mathbf{r}-\mathbf{r}_{A/B})=\sqrt{\frac{1}{\pi l^{2}}}\exp[-r^{2}/(2l^{2}%
)]$, where $l=\sqrt{\hslash/(m\omega_{0})}$ with $\omega_{0}=\sqrt{3V_{0}%
E_{r}^{L}}/\hslash$, of which the recoil energy reads $E_{r}^{L}=\hbar
^{2}k_{L}^{2}/(2m)$ and $m$ is a single particle mass. Here $A$, $B$ denote
two-sublattice, $\alpha=1,$ $2$ denote the index of the layer and
$\sigma=\uparrow,$ $\downarrow$ denote (pseudo-)spin degree of freedom,
respectively. Then the two-component fermions in 2D bilayer honeycomb optical
lattice has a Hamiltonian as\cite{Haldane}
\begin{align}
\hat{H}_{\mathrm{bi}} &  =-t\sum \limits_{\left \langle {i,j}\right \rangle
,\alpha,\sigma}\hat{c}_{i,\alpha,\sigma}^{\dagger}\hat{c}_{j,\alpha,\sigma
}-t^{\prime}\sum \limits_{\left \langle \left \langle {i,j}\right \rangle
\right \rangle ,\alpha,\sigma}e^{\mathrm{i}\phi_{ij}}\hat{c}_{i,\alpha,\sigma
}^{\dagger}\hat{c}_{j,\alpha,\sigma}\label{hop}\\
&  -t_{\perp}\sum \limits_{{i},\sigma}\hat{c}_{i,1,\sigma}^{\dagger}\hat
{c}_{i,2,\sigma}+h.c.,\nonumber
\end{align}
where $\hat{c}_{i,\alpha,\sigma}^{\dagger}$ ($\hat{c}_{i,\alpha,\sigma}$)
represents fermion creation (annihilation) operators at site $i$ of
layer-$\alpha$ ($\alpha=1$ or $2$) for spin $\sigma$ ($\uparrow$ or
$\downarrow$). $t$ ($t^{\prime}$) is the real nearest (next nearest) neighbor
hopping amplitude, and $t_{\perp}$ is the interlayer coupling which is much
smaller than $t$, i.e., $t_{\perp}\ll t$. $\left \langle {i,j}\right \rangle $,
$\left \langle \left \langle {i,j}\right \rangle \right \rangle $ denote the
nearest neighbor and the next nearest neighbor links, respectively.\ The next
nearest neighbor hopping term has a complex phase $\phi_{ij}=\pm \frac{\pi}{2}%
$, where the positive phase is set clockwise. To design a complex phase of the
next nearest neighbor hopping for a two-component fermions generated by the
gauge field on the optical lattice, we may apply a Raman laser beams in XY
plane with spacial-dependent Rabi frequencies as $\Omega_{0}\sin(\tilde{k}%
_{L}x+\frac{\pi}{4})e^{\mathrm{i}y\tilde{k}}$ and $\Omega_{0}\cos(\tilde
{k}_{L}x+\frac{\pi}{4})e^{-\mathrm{i}y\tilde{k}}$ ($\tilde{k}_{L}=\frac{2\pi
}{3a}$) where $a$ denotes the length between nearest neighbour lattice
sites.\textbf{ }Then we get a laser-field-generated effective gauge field on
this honeycomb optical lattice similar to that proposed for the monolayer
honeycomb optical lattice in Ref.\cite{shao}.

In addition, we apply a layer-dependent Zeeman field to polarize spin degree
of freedom by detuning Raman lasers only on the fermions in layer-1 as
\begin{equation}
\hat{H}_{\mathrm{Zeeman}}=h\sum \limits_{i,\sigma,\sigma^{\prime}}\hat
{c}_{i,1,\sigma}^{\dagger}\sigma_{\sigma,\sigma^{\prime}}^{z}\hat
{c}_{i,1,\sigma^{\prime}}.
\end{equation}
For big enough Zeeman field $h$, this term eventually drives this fermion
model at $1/4$ filling in layer-1 to a $\mathcal{C}=\pm1$ topological
insulator with fixed chemical potential $\mu_{1}=-h$. Such layer-dependent
Zeeman field can be realized by the a modulated laser wave along z-direction,
of which the wave vector is $k_{L}^{z}$ but has an additional phase shift to
the laser beams that generate the optical lattice, $\Delta \phi=2\pi
-\arccos \left[  \sqrt{V_{L}/\left(  4V_{s}\right)  }\right]  $. For this case,
we always get a zero Zeeman field at layer-2 at positions with a minimum
potential along z-direction $k_{L}^{z}z=2\pi n-\arccos \left[  \sqrt
{V_{L}/\left(  4V_{s}\right)  }\right]  $ but a finite Zeeman field at layer-1
at positions with a minimum potential along z-direction $k_{L}^{z}z=2\pi
n+\arccos \left[  \sqrt{V_{L}/\left(  4V_{s}\right)  }\right]  $.

Furthermore, we consider a strong interaction via Feshbach resonance
technique\cite{fesh,fesh1} and contact interaction%
\begin{equation}
\hat{H}_{U}=-U\sum \limits_{i,\alpha=1,2}\hat{n}_{i,\alpha,\uparrow}\hat
{n}_{i,\alpha,\downarrow}%
\end{equation}
where $U>0$ is the on-site attractive interaction strength given by integrals
over the Wannier function around site $A$ or $B$ that reads
\begin{align}
U  &  =-g\int d\boldsymbol{r}\left \vert w(\mathbf{r}-\mathbf{r}_{A/B}%
)\right \vert ^{4}\label{U}\\
&  \simeq2\sqrt{3}\{[\ln(k_{L}^{2}b^{2}/4)]^{-1}+\Delta \mathcal{B}%
/(\mathcal{B}-\mathcal{B}_{0})\} \sqrt{V_{0}E_{r}^{L}}\nonumber
\end{align}
where $g$ is coupling constant in two dimension\cite{morgan}, $b$ is radius of
hard-sphere potential, $\mathcal{B}$ is magnetic field, $\mathcal{B}_{0}$ is
resonance magnetic field and $\Delta \mathcal{B}$ is the width of the
resonance, respectively.

Finally we get an interacting two-component fermions system in 2D bilayer
honeycomb optical lattice of the Haldane model with layer-dependent Zeeman
field as\cite{Haldane,he1,he2}
\begin{equation}
\hat{H}=\hat{H}_{\mathrm{bi}}+\hat{H}_{\mathrm{Zeeman}}+\hat{H}_{U}+\hat
{H}_{c}. \label{haldane}%
\end{equation}
where
\begin{equation}
\hat{H}_{c}=-\mu_{1}\sum \limits_{{i},\sigma}\hat{c}_{i,1,\sigma}^{\dagger}%
\hat{c}_{i,1,\sigma}-\mu_{2}\sum \limits_{{i},\sigma}\hat{c}_{i,2,\sigma
}^{\dagger}\hat{c}_{i,2,\sigma}%
\end{equation}
with $\mu_{1}$ and $\mu_{2}$ denoting the chemical potentials in layer-1 and
layer-2, respectively.

\section{Mean field approach and global phase diagram}

\begin{figure}[ptb]
\includegraphics[width=0.5\textwidth]{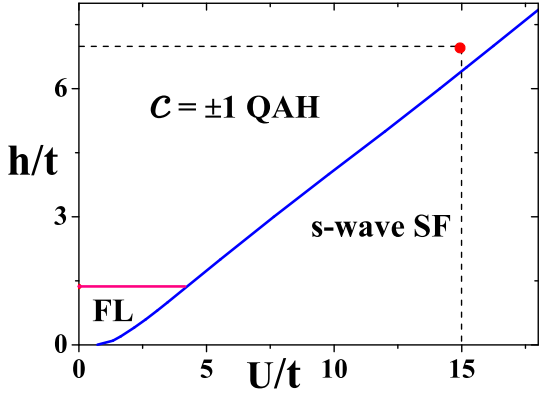}\caption{The phase diagram of
fermions (filling factor $n_{f}=1/4$) with attractive interaction on layer-1
honeycomb optical lattice for the case of $t^{\prime}=0.05t$. There exist
three quantum phases: $\mathcal{C}=\pm1$ QAH state, s-wave SF state and Fermi
liquid state (FL). In this paper we only consider the $\mathcal{C}=\pm1$ QAH
state of $h=7t$ which is marked by the red dot.}%
\label{ph}%
\end{figure}

Considering a tiny interlayer coupling $t_{\perp}$, we may use the mean field
approach separately for each layer. When increasing the interaction strength,
the fermionic system described in Eq.(\ref{haldane}) is unstable against
superfluid (SF) orders that are described by $\triangle_{1,2}$ for s-wave
pairing order parameters as $\Delta_{1}=\left \langle \hat{c}_{i,1,\downarrow
}\hat{c}_{i,1,\uparrow}\right \rangle $ and $\Delta_{2}=\left \langle \hat
{c}_{i,2,\downarrow}\hat{c}_{i,2,\uparrow}\right \rangle $. Due to the
layer-dependent Zeeman field, the symmetry between two layers is broken
($\Delta_{1}\neq \Delta_{2}$). In this section, we give the mean field
calculations. Due to tiny inter-layer coupling, we may do calculations of the
models for each layer separately.

Firstly, we consider the interacting Haldane model on layer-1 as%
\begin{align}
\hat{H}_{1}  &  =-t\sum \limits_{\left \langle {i,j}\right \rangle ,\sigma}%
\hat{c}_{i,1,\sigma}^{\dagger}\hat{c}_{j,1,\sigma}-t^{\prime}\sum
\limits_{\left \langle \left \langle {i,j}\right \rangle \right \rangle ,\sigma
}e^{\mathrm{i}\phi_{ij}}\hat{c}_{i,1,\sigma}^{\dagger}\hat{c}_{j,1,\sigma}\\
&  -\mu \sum \limits_{{i},\sigma}\hat{c}_{i,1,\sigma}^{\dagger}\hat
{c}_{i,1,\sigma}+h\sum \limits_{i,\sigma,\sigma^{\prime}}\hat{c}_{i,1,\sigma
}^{\dagger}\sigma_{\sigma,\sigma^{\prime}}^{z}\hat{c}_{i,1,\sigma^{\prime}%
}\nonumber \\
&  -U\sum \limits_{i}\hat{n}_{i,1,\uparrow}\hat{n}_{i,1,\downarrow
}+h.c..\nonumber
\end{align}
When there are a finite s-wave pairing order parameters of layer-1 $\Delta
_{1}=\left \langle \hat{c}_{i,1,\uparrow}\hat{c}_{i,1,\downarrow}\right \rangle
$, we get the effective Hamiltonian as
\begin{align}
\hat{H}_{1,\mathrm{eff}}  &  =-t\sum \limits_{\left \langle {i,j}\right \rangle
,\sigma}\hat{c}_{i,1,\sigma}^{\dagger}\hat{c}_{j,1,\sigma}-t^{\prime}%
\sum \limits_{\left \langle \left \langle {i,j}\right \rangle \right \rangle
,\sigma}e^{\mathrm{i}\phi_{ij}}\hat{c}_{i,1,\sigma}^{\dagger}\hat
{c}_{j,1,\sigma}\\
&  +h\sum \limits_{i,\sigma,\sigma^{\prime}}\hat{c}_{i,1,\sigma}^{\dagger
}\sigma_{\sigma,\sigma^{\prime}}^{z}\hat{c}_{i,1,\sigma^{\prime}}-U\sum
_{i}\Delta_{1}\hat{c}_{i\mathbf{,}1,\downarrow}\hat{c}_{i\mathbf{,}1,\uparrow
}\nonumber \\
&  -\mu \sum \limits_{{i},\sigma}\hat{c}_{i,1,\sigma}^{\dagger}\hat
{c}_{i,1,\sigma}+h.c..\nonumber
\end{align}
The energy spectrums of the fermions in layer-1 are given by
\begin{align}
\left(  E_{1,k}\right)  _{1}  &  =h+e_{_{1,k}},\text{ }\left(  E_{1,k}\right)
_{2}=h-e_{_{1,k}}\\
\left(  E_{1,k}\right)  _{3}  &  =h+e_{_{2,k}},\text{ }\left(  E_{1,k}\right)
_{4}=h-e_{_{2,k}}\nonumber
\end{align}
where
\begin{align}
e_{1,k}  &  =\sqrt{\mu^{2}+(U\Delta)^{2}+\gamma_{k}^{2}+|\xi_{k}|^{2}%
+2f},\nonumber \\
e_{2,k}  &  =\sqrt{\mu^{2}+(U\Delta)^{2}+\gamma_{k}^{2}+|\xi_{k}|^{2}-2f}.
\end{align}
The functions $|\xi_{k}|$, $\gamma_{k}$, $f$ are
\begin{align}
|\xi_{k}|  &  =t\sqrt{3+2\cos(\sqrt{3}k_{y})+4\cos(3k_{x}/2)\cos(\sqrt{3}%
k_{y}/2)},\nonumber \\
\gamma_{k}  &  =-t^{\prime}[4\cos(3k_{x}/2)\sin(\sqrt{3}k_{y}/2)-2\sin \sqrt
{3}k_{y}],\nonumber \\
f  &  =2\sqrt{\gamma_{k}^{2}[\mu^{2}+(U\Delta)^{2}]+|\xi_{k}|^{2}\mu^{2}}.
\end{align}

By minimizing the ground state energy we arrive at the following
self-consistent equations
\begin{align}
\frac{1}{U}  &  =\frac{1}{4N}[-\sum_{(E_{1,k})_{1}<0}\frac{1+\gamma_{k}^{2}%
/f}{e_{_{1,k}}}\tanh(-\frac{\left(  E_{1,k}\right)  _{1}}{2T})\\
&  +\sum_{(E_{1,k})_{2}<0}\frac{1+\gamma_{k}^{2}/f}{e_{_{1,k}}}\tanh
(-\frac{\left(  E_{1,k}\right)  _{2}}{2T})\nonumber \\
&  -\sum_{(E_{1,k})_{3}<0}\frac{1-\gamma_{k}^{2}/f}{e_{_{2,k}}}\tanh
(-\frac{\left(  E_{1,k}\right)  _{3}}{2T})\nonumber \\
&  +\sum_{(E_{1,k})_{4}<0}\frac{1-\gamma_{k}^{2}/f}{e_{_{2},k}}\tanh
(-\frac{\left(  E_{1,k}\right)  _{4}}{2T})],\nonumber
\end{align}
and%
\begin{align}
n  &  =1+\frac{\mu}{2N}[-\sum_{(E_{1,k})_{1}<0}\frac{1+(\gamma_{k}^{2}%
+|\xi_{k}|^{2})/f}{e_{_{1,k}}}\tanh(-\frac{\left(  E_{1,k}\right)  _{1}}%
{2T})\\
&  +\sum_{(E_{1,k})_{2}<0}\frac{1+(\gamma_{k}^{2}+|\xi_{k}|^{2})/f}{e_{_{1,k}%
}}\tanh(-\frac{\left(  E_{1,k}\right)  _{2}}{2T})\nonumber \\
&  -\sum_{(E_{1,k})_{3}<0}\frac{1-(\gamma_{k}^{2}+|\xi_{k}|^{2})/f}{e_{_{2,k}%
}}\tanh(-\frac{\left(  E_{1,k}\right)  _{3}}{2T})\nonumber \\
&  +\sum_{(E_{1,k})_{4}<0}\frac{1-(\gamma_{k}^{2}+|\xi_{k}|^{2})/f}{e_{_{2,k}%
}}\tanh(-\frac{\left(  E_{1,k}\right)  _{4}}{2T})],\nonumber
\end{align}
where $n$ is fermion density and $N$ is the number of primitive cells. By this
approach we obtain the phase diagram given in Fig. 4.

\begin{figure}[ptb]
\includegraphics[width=0.5\textwidth]{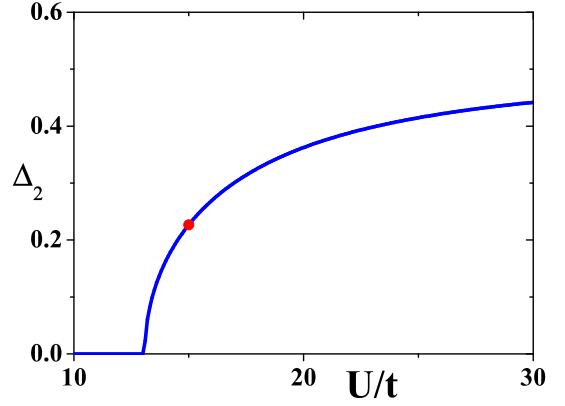}\caption{SF order parameter of
the fermion with attractive interaction on layer-2 honeycomb optical lattice
for the case of $t^{\prime}=0.05t$. There exist two quantum phases: s-wave SF
state and metal state. In this paper we only consider the s-wave SF state of
$\mu_{2}=-7t$ which is marked by the red point.}%
\label{sfo}%
\end{figure}

Next we consider the interacting Haldane model on layer-2 as%
\begin{align}
\hat{H}_{2}  &  =-t\sum \limits_{\left \langle {i,j}\right \rangle ,\sigma}%
\hat{c}_{i,2,\sigma}^{\dagger}\hat{c}_{j,2,\sigma}-t^{\prime}\sum
\limits_{\left \langle \left \langle {i,j}\right \rangle \right \rangle ,\sigma
}e^{i\phi_{ij}}\hat{c}_{i,2,\sigma}^{\dagger}\hat{c}_{j,2,\sigma}\nonumber \\
&  -\mu \sum \limits_{{i},\sigma}\hat{c}_{i,2,\sigma}^{\dagger}\hat
{c}_{i,2,\sigma}-U\sum \limits_{i}\hat{n}_{2,i\uparrow}\hat{n}_{2,i\downarrow}.
\end{align}
of which the chemical potential is set to be equal to that of layer-1.
Considering a finite s-wave pairing order parameters of layer-2 $\Delta
_{2}=\left \langle \hat{c}_{i,2,\uparrow}\hat{c}_{i,2,\downarrow}\right \rangle
$, we get the effective Hamiltonian as
\begin{align}
\hat{H}_{2,\mathrm{eff}}  &  =-t\sum \limits_{\left \langle {i,j}\right \rangle
,\sigma}\hat{c}_{i,2,\sigma}^{\dagger}\hat{c}_{j,2,\sigma}-t^{\prime}%
\sum \limits_{\left \langle \left \langle {i,j}\right \rangle \right \rangle
,\sigma}e^{\mathrm{i}\phi_{ij}}\hat{c}_{i,2,\sigma}^{\dagger}\hat
{c}_{j,2,\sigma}\nonumber \\
&  -U\sum_{i}\Delta_{2}\hat{c}_{i\mathbf{,}2,\downarrow}\hat{c}_{i\mathbf{,}%
2,\uparrow}-\mu \sum \limits_{{i},\sigma}\hat{c}_{i,2,\sigma}^{\dagger}\hat
{c}_{i,2,\sigma}+h.c..
\end{align}
The energy spectrums of the fermions in layer-1 are given by
\[
\left(  E_{2,k}\right)  _{1}=\left(  E_{2,k}\right)  _{2}=e_{_{1,k}},\left(
E_{2,k}\right)  _{3}=\left(  E_{2,k}\right)  _{4}=e_{_{2,k}}.
\]
Similarly, we get self-consistent equations as%
\begin{align}
\frac{1}{U}  &  =\frac{1}{4N}[\sum_{k}\frac{1+\gamma_{k}^{2}/f}{e_{_{1,k}}%
}\tanh(\frac{e_{_{1,k}}}{2T})\\
&  +\sum_{k}\frac{1-\gamma_{k}^{2}/f}{e_{_{2,k}}}\tanh(\frac{e_{_{2,k}}}%
{2T})]\nonumber
\end{align}
and%
\begin{align}
n  &  =1+\frac{\mu}{2N}[\sum_{k}\frac{1+(\gamma_{k}^{2}+|\xi_{k}|^{2}%
)/f}{e_{_{1,k}}}\tanh(\frac{e_{_{1,k}}}{2T})\\
&  \sum_{k}\frac{1-(\gamma_{k}^{2}+|\xi_{k}|^{2})/f}{e_{_{2,k}}}\tanh
(\frac{e_{_{2,k}}}{2T})].\nonumber
\end{align}
By this approach we obtain the SF pairing order parameter given in
Fig.\ref{sfo}.

Now the total effective Hamiltonian is given by
\begin{equation}
\hat{H}_{\mathrm{eff}}=\hat{H}_{\mathrm{bi}}+\hat{H}_{\mathrm{Zeeman}}+\hat
{H}_{SF}+\hat{H}_{c},
\end{equation}
where
\begin{equation}
\hat{H}_{SF}=-U\sum_{i,\alpha}\Delta_{\alpha}(\hat{c}_{i\mathbf{,}%
\alpha,\downarrow}\hat{c}_{i\mathbf{,}\alpha,\uparrow}+\hat{c}_{i\mathbf{,}%
\alpha,\uparrow}^{\dagger}\hat{c}_{i\mathbf{,}\alpha,\downarrow}^{\dagger}).
\end{equation}
Thus we get four self-consistent equations of the fermions in both layers for
$\Delta_{1,2}$ and $\mu_{1,2}$ by minimizing the ground state energy. We fix
the fermion filling factor $n_{f}=1/4$ in layer-1. A key point here is to keep
the following chemical potential condition:
\begin{equation}
\mu_{1}=\mu_{2}%
\end{equation}
which guarantees $1/4$ filling factor for the fermions in layer-1. To keep the
above condition, we must tune the chemical potential $\mu_{2}$\ in layer-2 by
manipulating the interaction $U$ to synchronize with $\mu_{1}$ in layer-1.

From Fig.\ref{ph}, one can see that for $U\neq0$, there exist three quantum
phases in layer-1: the Fermi liquid (FL), the topological insulator of
$\mathcal{C}=\pm1$ TKNN number with quantized anomalous Hall (QAH) effect (we
also call it $\mathcal{C}=\pm1$ QAH state) and s-wave superfluid
(SF)\cite{thou}. For free fermions, $U=0,$ there exists a critical point for
Zeeman field: $h_{c}\simeq1.37t$. For the case of large Zeeman field,
$h>h_{c}$, the ground state is $\mathcal{C}=\pm1$ topological insulator with
chiral edge states; For the case of smaller Zeeman field, $h<h_{c}$, the
ground state is a FL state. When considering the interaction term, there may
exist s-wave SF state, and the quantum phase transition from $\mathcal{C}%
=\pm1$ QAH state to SF state is the first order. In this paper we consider the
$\mathcal{C}=\pm1$ QAH state with $h=7t$. For this case, due to big imbalanced
Fermi gaps of different spin components, the attracting interaction is
irrelevant. The ground state is $\mathcal{C}=\pm1$ QAH state until the
interaction strength is larger than $16.3t$.

For layer-2, the situation is much different due to zero Zeeman field.
Fig.\ref{sfo} shows the SF pairing order parameter $\Delta_{2}$ of layer-2 for
the case of $t^{\prime}=0.05t,$ $\mu_{2}=-7t$. For the weak interaction case,
the ground state is FL; for the large interaction case, when $U>13.0t$, the
ground state becomes SF with trivial topological properties. Just for this
reason, we cannot get a $\mathcal{C}=\pm1$ topological SF in a monolayer
system, and thus have to turn to a bilayer system.

Now, we consider the model with the parameters $U=15t,$ $h=7t,$ $t^{\prime
}=0.05t$ (marked by the red spots in Fig.\ref{ph} and Fig.\ref{sfo},
respectively), at which we derived $\Delta_{1}=0,$ $\Delta_{2}=0.227$,
$\mu_{1}=\mu_{2}=-h=-7t$ at zero temperature. Thus we have a $\mathcal{C}%
=\pm1$ QAH in layer-1 and an s-wave SF state in layer-2. Now the total
Hamiltonian is given by
\begin{align}
\hat{H}_{\mathrm{eff}}  &  =-t\sum \limits_{\left \langle {i,j}\right \rangle
,\alpha,\sigma}\hat{c}_{i,\alpha,\sigma}^{\dagger}\hat{c}_{j,\alpha,\sigma
}-t^{\prime}\sum \limits_{\left \langle \left \langle {i,j}\right \rangle
\right \rangle ,\alpha,\sigma}e^{\mathrm{i}\phi_{ij}}\hat{c}_{i,\alpha,\sigma
}^{\dagger}\hat{c}_{j,\alpha,\sigma}\\
&  -t_{\perp}\sum \limits_{{i},\sigma}(\hat{c}_{i,1,\sigma}^{\dagger}\hat
{c}_{i,2,\sigma}+h.c.)+h\sum \limits_{i,\sigma,\sigma^{\prime}}\hat
{c}_{i,1,\sigma}^{\dagger}\sigma_{\sigma,\sigma^{\prime}}^{z}\hat
{c}_{i,1,\sigma^{\prime}}\nonumber \\
&  +h\sum \limits_{{i},\sigma}\hat{c}_{{i},1,\sigma}^{\dagger}\hat{c}%
_{{i},1,\sigma}+h\sum \limits_{{i},\sigma}\hat{c}_{{i},2,\sigma}^{\dagger}%
\hat{c}_{{i},2,\sigma}\nonumber \\
&  -U\sum_{i}\Delta_{2}(\hat{c}_{i\mathbf{,}2,\downarrow}\hat{c}%
_{i\mathbf{,}2,\uparrow}+\hat{c}_{i\mathbf{,}2,\uparrow}^{\dagger}\hat
{c}_{i\mathbf{,}2,\downarrow}^{\dagger}).\nonumber
\end{align}
At high temperature, the SF pairing order disappears. The transition
temperature of SF order in layer-2 is $k_{B}T_{c}\simeq2.25t$.

\section{$\mathcal{C}=\pm1$ topological SF\textrm{ }due to proximity effect of
SF order in layer-2 on $\mathcal{C}=\pm1$ QAH in layer-1}

\begin{figure}[ptb]
\includegraphics[width=0.5\textwidth]{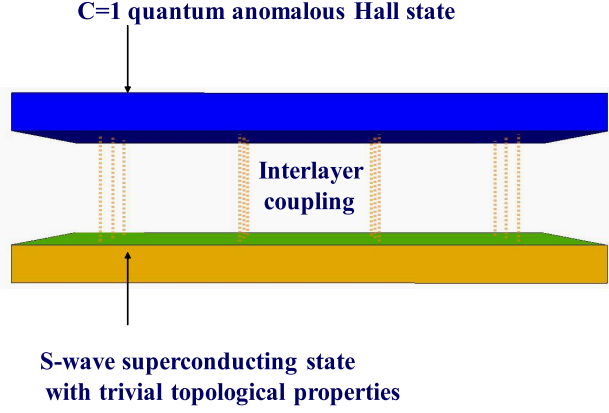}\caption{The illustration of
$\mathcal{C}=\pm1$ topological SF due to proximity effect between SF order in
layer-2 and $\mathcal{C}=\pm1$ QAH in layer-1. }%
\label{twol}%
\end{figure}

We use the purterbative theory to calculate the proximity effect of the SF
order in layer-2 on the $\mathcal{C}=\pm1$ QAH in layer-1. See the
illustration in Fig.\ref{twol}. The Hamiltonian has a form as
\begin{equation}
\hat{H}_{\mathrm{eff}}=\hat{H_{0}}+\hat{H}_{I}%
\end{equation}
in which%
\begin{align}
\hat{H_{0}}  &  =-t\sum \limits_{\left \langle {i,j}\right \rangle ,\alpha
,\sigma}\hat{c}_{i,\alpha,\sigma}^{\dagger}\hat{c}_{j,\alpha,\sigma}%
-t^{\prime}\sum \limits_{\left \langle \left \langle {i,j}\right \rangle
\right \rangle ,\alpha,\sigma}e^{\mathrm{i}\phi_{ij}}\hat{c}_{i,\alpha,\sigma
}^{\dagger}\hat{c}_{j,\alpha,\sigma}\\
&  +h\sum \limits_{i,\sigma,\sigma^{\prime}}\hat{c}_{i,1,\sigma}^{\dagger
}\sigma_{\sigma,\sigma^{\prime}}^{z}\hat{c}_{i,1,\sigma^{\prime}}-\mu
\sum \limits_{{i},\sigma}\hat{c}_{i,1,\sigma}^{\dagger}\hat{c}_{i,1,\sigma
}\nonumber \\
&  -U\sum_{i}\Delta_{2,i}(\hat{c}_{i\mathbf{,}2,\downarrow}\hat{c}%
_{i\mathbf{,}2,\uparrow}+\hat{c}_{i\mathbf{,}2,\uparrow}^{\dagger}\hat
{c}_{i\mathbf{,}2,\downarrow}^{\dagger})+h.c.\nonumber \\
&  -\mu \sum \limits_{{i},\sigma}\hat{c}_{i,2,\sigma}^{\dagger}\hat
{c}_{i,2,\sigma}\nonumber
\end{align}
is the unperturbation term, and due to $t_{\perp}\ll U\Delta_{2}$ and the
interlayer coupling
\begin{equation}
\hat{H}_{I}=-t_{\perp}\sum \limits_{{i},\sigma}\left(  \hat{c}_{i,1,\sigma
}^{\dagger}\hat{c}_{i,2,\sigma}+h.c.\right)
\end{equation}
is the small perturbation term.

In the purterbative theory, we firstly use the path-integral representation
\begin{equation}
Z=\int[dc_{1,\sigma}^{\ast}dc_{1,\sigma}][dc_{2,\sigma}^{\ast}dc_{2,\sigma
}]e^{-S_{0}-S^{\prime}}%
\end{equation}
by replacing electronic operators $\hat{c}_{i,\sigma}^{\dagger}$ and $\hat
{c}_{j,\sigma}$ to Grassmann variables $c_{i,\sigma}^{\ast}$ and $c_{j,\sigma
}$. $S_{0}$ is the action as
\begin{equation}
S_{0}=\int d\tau \mathcal{L}_{0}%
\end{equation}
and the Lagrangian in terms of Grassmann variables $c_{i,\sigma}^{\ast}$ and
$c_{i,\sigma}$ is then obtained as
\begin{equation}
\mathcal{L}_{0}=\sum_{i,\alpha,\sigma}c_{i,\alpha,\sigma}^{\ast}\partial
_{\tau}c_{i,\alpha,\sigma}+H_{0}(c^{\ast},c),\nonumber
\end{equation}
where $H_{0}(c^{\ast},c)$ is obtained by replacing operators in $\hat{H}_{0}$
with Grassman variables. $S^{\prime}$ is the action as
\begin{equation}
S^{\prime}=\int d\tau \mathcal{L}^{\prime}%
\end{equation}
where
\begin{equation}
\mathcal{L}^{\prime}=-t_{\perp}\sum \limits_{{i},\sigma}\left(  c_{i,1,\sigma
}^{\ast}c_{i,2,\sigma}+h.c.\right)  .
\end{equation}

Now we integrate $c_{2,i,\sigma}^{\ast}$ $c_{2,i,\sigma}$ and get
\begin{equation}
Z=\int[dc_{1,\sigma}^{\ast}dc_{1,\sigma}]e^{-(S_{1})_{0}-S_{\mathrm{eff}%
}^{\prime}}%
\end{equation}
where
\begin{align}
(S_{1})_{0}  &  =\int d\tau \lbrack \sum_{i,\sigma}c_{i,1,\sigma}^{\ast}%
\partial_{\tau}c_{i,1,\sigma}-t\sum \limits_{\left \langle ij\right \rangle
,\sigma}c_{i,1,\sigma}^{\ast}c_{j,1,\sigma}\\
&  -\mu_{1}\sum \limits_{{i,}\sigma}c_{i,1,\sigma}^{\ast}c_{i,1,\sigma
}-t^{\prime}\sum \limits_{\left \langle \left \langle {i,j}\right \rangle
\right \rangle ,\sigma}e^{\mathrm{i}\phi_{ij}}c_{i,1,\sigma}^{\ast
}c_{j,1,\sigma}\nonumber \\
&  +h\sum \limits_{i,\sigma}c_{i,1,\sigma}^{\ast}\sigma^{z}c_{i,1,\sigma
}+h.c.]\nonumber
\end{align}
and
\begin{equation}
S_{\mathrm{eff}}^{\prime}=-\ln \left \langle e^{-S^{\prime}}\right \rangle _{2}%
\end{equation}
where $\left \langle e^{-S^{\prime}}\right \rangle _{2}$ $=\int[dc_{2}^{\ast
}dc_{2}]e^{-\left(  S_{2}\right)  _{0}-S^{\prime}}.$ Thus we have
\begin{align}
\left \langle e^{-S^{\prime}}\right \rangle _{2}  &  =\int[dc_{2,\sigma}^{\ast
}dc_{2,\sigma}]e^{-\left(  S_{2}\right)  _{0}-S^{\prime}}\\
&  \simeq e^{\left[  \left \langle -S^{\prime}\right \rangle _{2}+\frac{1}%
{2}\left(  \left \langle S^{\prime2}\right \rangle _{2}-\left \langle -S^{\prime
}\right \rangle _{2}^{2}\right)  +\cdot \cdot \cdot \right]  .}\nonumber
\end{align}
Due to $\left \langle S^{\prime}\right \rangle _{2}=0$ we derive
\begin{align}
S_{\mathrm{eff}}^{\prime}  &  =-\ln \left \langle e^{-S^{\prime}}\right \rangle
_{2}\simeq-\frac{1}{2}\left \langle \left(  S^{\prime}\right)  ^{2}%
\right \rangle _{2}\\
&  =\int d\tau \lbrack \sum_{i}\Delta_{1,\mathrm{induce}}c_{i\mathbf{,}%
1,\downarrow}c_{i\mathbf{,}1,\uparrow}+h.c.]\nonumber
\end{align}
where
\begin{equation}
\Delta_{1,\mathrm{induce}}=-\frac{\left(  t_{\perp}\right)  ^{2}}%
{U\Delta_{i,2}}.
\end{equation}
To derive this result we have used the following equation,%
\begin{align}
\left \langle \left(  S^{\prime}\right)  ^{2}\right \rangle _{2}  &
=\left \langle (t_{\perp}\sum \limits_{i,\sigma}c_{i,1,\sigma}^{\ast
}c_{i,2,\sigma}+h.c.)^{2}\right \rangle _{2}\\
&  =2t_{\perp}^{2}\sum \limits_{i}[c_{i\mathbf{,}1,\downarrow}c_{i\mathbf{,}%
1,\uparrow}\left \langle c_{i\mathbf{,}2,\uparrow}^{\ast}c_{i\mathbf{,}%
2,\downarrow}^{\ast}\right \rangle _{2}\nonumber \\
&  +c_{i\mathbf{,}1,\uparrow}^{\ast}c_{i\mathbf{,}1,\downarrow}^{\ast
}\left \langle c_{i\mathbf{,}2,\downarrow}c_{i\mathbf{,}2,\uparrow
}\right \rangle _{2}]\nonumber \\
&  =\frac{2t_{\perp}^{2}}{U\Delta_{2,i}}\sum \limits_{i}\left[  c_{i\mathbf{,}%
1,\downarrow}c_{i\mathbf{,}1,\uparrow}+c_{i\mathbf{,}1,\uparrow}^{\ast
}c_{i\mathbf{,}1,\downarrow}^{\ast}\right] \nonumber
\end{align}
where
\begin{equation}
\left \langle c_{i\mathbf{,}2,\downarrow}c_{i\mathbf{,}2,\uparrow}\right \rangle
_{2}=\left \langle c_{i\mathbf{,}2,\uparrow}^{\ast}c_{i\mathbf{,}2,\downarrow
}^{\ast}\right \rangle _{2}=\frac{1}{U\Delta_{i,2}}.
\end{equation}

That means although there is no SF pairing order parameter of fermions in
layer-1, the tiny interlayer hopping will lead to an induced SF order due to
the proximity effect. After integrating gapped fermions on layer-2, the low
energy effective model of such bilayer system is finally reduced into
\begin{align}
\hat{H}_{1,\mathrm{eff}}  &  =-t\sum \limits_{\left \langle {i,j}\right \rangle
,\sigma}\hat{c}_{i,1,\sigma}^{\dagger}\hat{c}_{j,1,\sigma}-t^{\prime}%
\sum \limits_{\left \langle \left \langle {i,j}\right \rangle \right \rangle
,\sigma}e^{\mathrm{i}\phi_{ij}}\hat{c}_{i,1,\sigma}^{\dagger}\hat
{c}_{j,1,\sigma}\label{model}\\
&  +h\sum \limits_{i,\sigma,\sigma^{\prime}}\hat{c}_{i,1,\sigma}^{\dagger
}\sigma_{\sigma,\sigma^{\prime}}^{z}\hat{c}_{i,1,\sigma^{\prime}}%
+h\sum \limits_{{i},\sigma}\hat{c}_{{i},1,\sigma}^{\dagger}\hat{c}%
_{{i},1,\sigma}\nonumber \\
&  +\sum_{i}\Delta_{1,\mathrm{induce}}\hat{c}_{i\mathbf{,}1,\downarrow}\hat
{c}_{i\mathbf{,}1,\uparrow}+h.c.\nonumber
\end{align}
where $\Delta_{1,\mathrm{induce}}$ is the induced SF order parameter of
fermions in layer-1 which is estimated by the perturbation approach as
$\Delta_{1,\mathrm{induce}}\simeq-\frac{\left(  t_{\perp}\right)  ^{2}%
}{U\Delta_{2}}$.\textbf{ }And\textbf{ }for a tiny induced SF order parameter,
we always get a really large energy gap of the fermions in the bulk, $\Delta
E\simeq0.52t,$ which protects the topological properties of the $\mathcal{C}%
=\pm1$ topological SF order.

Thus due to the proximity effect between $\mathcal{C}=\pm1$ QAH and s-wave SF,
the ground state is really a $\mathcal{C}=\pm1$ topological SF as
\begin{equation}
\mathcal{C}=\pm1\text{ \textrm{QAH}}+\text{\textrm{s-wave SF}}\rightarrow
\mathcal{C}=\pm1\text{ \textrm{TSF}}%
\end{equation}
of which the topological properties is similar to that of 2D chiral
$p_{x}+\mathrm{i}p_{y}$ wave SF\cite{qi,yu}.

\section{Topological properties}

In this section we will study its topological properties by calculating the
edge states and the zero modes on a $\pi$-flux (vortex with half quantized
"magnetic" flux).

\begin{figure}[ptb]
\includegraphics[width=0.5\textwidth]{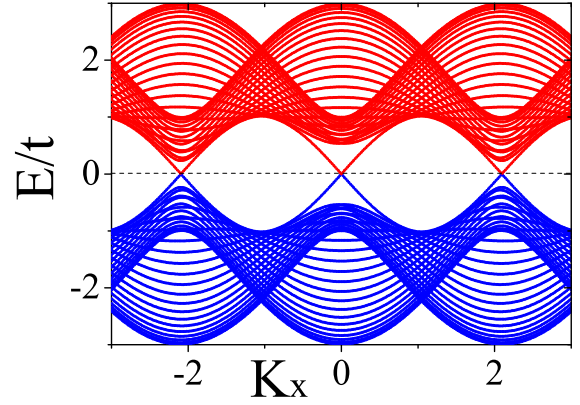}\caption{The armchair edge
state of $\mathcal{C}=\pm1$ QAH with induced SF order on layer-1 and related
parameters $U=15t,$ $h=7t,$ $\mu_{1}=-7t$, $t^{\prime}=0.05t$, $\Delta
_{1,\mathrm{induce}}=0.001$.}%
\label{edge}%
\end{figure}

In Fig.\ref{edge}, we show the gapless Majorana edge modes of this effective
model on a lattice with open boundary condition along y-direction (armchair
edge) and periodic boundary condition along x-direction.

\begin{figure}[ptb]
\includegraphics[width=0.57\textwidth]{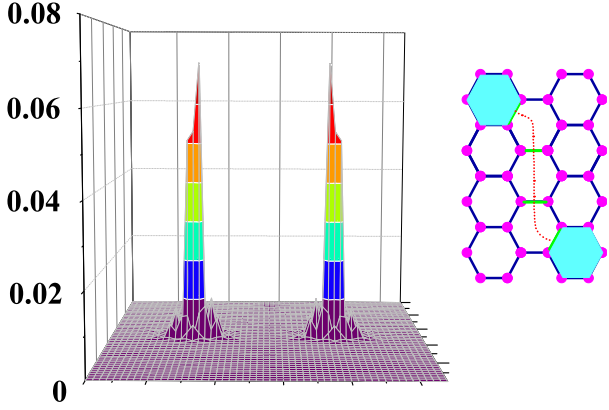}\caption{The particle density
of the zero modes of a pair of $\pi$-flux in $\mathcal{C}=\pm1$ QAH with
induced SF order parameter on a $36\times36$ lattice in layer-1. In the right
figure, there is a branch-cut (red dashed line) between two $\pi$-fluxes (cyan
plaquette) which changes the signs of the hopping terms on green links.}%
\label{zero}%
\end{figure}

In Fig.\ref{zero} we demonstrate the existence of a single zero-mode on a
$\pi$-flux obtained as a solution of the Bogoliubov-de Gennes equations
\cite{read,iva}. This is the Majorana zero energy mode and can be described by
a real fermion field $\gamma^{\dagger}=\int d{r}[u_{0}\psi^{\dagger}+v_{0}%
\psi]$ ($\gamma^{\dagger}=\gamma$)\cite{read}. When two $\pi$-fluxes are fused
together (taken to the same plaquette in the honeycomb optical lattice), the
result contains more than one quasiparticle due to the Ising fusion rule
\cite{mr},
\begin{equation}
\sigma \times \sigma=I+\psi. \label{fusion}%
\end{equation}
These results imply that the $\pi$-flux is a non-Abelian anyon (people also
call it Ising anyon) obeying non-Abelian statistics as that in chiral
$p_{x}+\mathrm{i}p_{y}$ wave SF. The topological properties of the
$\mathcal{C}=\pm1$ topological SF order are similar to those of chiral
$p_{x}+\mathrm{i}p_{y}$ superconductors with $\mathcal{C}=\pm1$ topological invariable.

\section{Phase stiffness of $\mathcal{C}=\pm1$ topological SF order}

In this section, by using the random-phase-approximation, we calculate the
phase stiffness of $\mathcal{C}=\pm1$ topological SF order which is determined
by the phase fluctuations in layer-2. The Hamiltonian of fermions in layer-2
is
\begin{align}
\hat{H}_{2,\mathrm{eff}}  &  =-t\sum \limits_{\left \langle {i,j}\right \rangle
,\sigma}\hat{c}_{i,2,\sigma}^{\dagger}\hat{c}_{j,2,\sigma}-t^{\prime}%
\sum \limits_{\left \langle \left \langle {i,j}\right \rangle \right \rangle
,\sigma}e^{\mathrm{i}\phi_{ij}}\hat{c}_{i,2,\sigma}^{\dagger}\hat
{c}_{j,2,\sigma}\nonumber \\
&  -\mu \sum \limits_{{i},\sigma}\hat{c}_{i,2,\sigma}^{\dagger}\hat
{c}_{i,2,\sigma}-U\sum \limits_{i}\hat{n}_{i,2,\uparrow}\hat{n}_{i,2,\downarrow
}+h.c..
\end{align}
Thus in path-integral representation the partition function\ is given by%
\begin{equation}
Z=\int[dc_{2,\sigma}^{\ast}dc_{2,\sigma}]e^{-S_{2}}%
\end{equation}
by replacing electronic operators $\hat{c}_{i,2,\sigma}^{\dagger}$ and
$\hat{c}_{j,2,\sigma}$ to Grassmann variables $c_{i,2,\sigma}^{\ast}$ and
$c_{j,2,\sigma}$. The effective action $S_{2}$ is
\begin{equation}
S_{2}=\int d\tau \mathcal{L}_{2}%
\end{equation}
and the Lagrangian in terms of Grassmann variables is then obtained as
\begin{equation}
\mathcal{L}_{2}=\sum_{i,\sigma}c_{i,2,\sigma}^{\ast}\partial_{\tau
}c_{i,2,\sigma}+H_{2,\mathrm{eff}}(c_{2,\sigma}^{\ast},c_{2,\sigma}),
\end{equation}
where $H_{2,\mathrm{eff}}(c_{2,\sigma}^{\ast},c_{2,\sigma})$ is obtained by
replacing the operators in $\hat{H}_{2,\mathrm{eff}}$ with Grassman variables.

Considering the s-wave pairing order parameter of layer-2 $\Delta
_{2}=\left \langle \hat{c}_{2,i,\uparrow}\hat{c}_{2,i,\downarrow}\right \rangle
$, we get the partition function as
\begin{equation}
Z=\int[dc_{2,\sigma}^{\ast}dc_{2,\sigma}][d\Delta^{\ast}d\Delta]e^{-S},
\end{equation}
where the action of fermions in layer-2 is
\begin{align}
S  &  =\sum \limits_{\omega_{m},\mathbf{k}}(\mathrm{i}\omega_{m}+\gamma
_{\mathbf{k}}-\mu)c_{\mathbf{k},\sigma}^{\ast}c_{\mathbf{k},\sigma}+2\beta \mu
N\\
&  +\sum \limits_{\omega_{m},\mathbf{k}}\xi_{\mathbf{k}}c_{A,\mathbf{k},\sigma
}^{\ast}c_{B,\mathbf{k},\sigma}+c.c.\nonumber \\
&  -\frac{1}{\sqrt{N}}\sum \limits_{k,q}(U\Delta^{\ast}(-q)c_{-\mathbf{k+q/2}%
,\uparrow}c_{\mathbf{k+q/2},\downarrow}\nonumber \\
&  +U\Delta(q)c_{\mathbf{k+q/2},\downarrow}^{\ast}c_{-\mathbf{k+q/2},\uparrow
}^{\ast})+2\beta N\sum \limits_{q}U\Delta^{\ast}(q)\Delta(q)\nonumber
\end{align}
with $\omega_{m}=(2n+1)\pi/\beta$ and $\omega_{l}=2n\pi/\beta$. We denote
$k=(\mathrm{i}\omega_{m},-\mathbf{k})$ and $q=(\mathrm{i}\omega_{l}%
,-\mathbf{q})$. After integrating over the fermionic field, the action turns
into%
\begin{equation}
S=2\beta N\sum \limits_{q}U\Delta^{\ast}(\mathbf{q})\Delta(\mathbf{q}%
)+2\beta \mu N-\sum \limits_{k,k^{\prime},q}\mathbf{Tr}\ln(G^{-1}),
\end{equation}
where\begin{widetext}
\[
G^{-1}=\left(
\begin{array}
[c]{cccc}%
(\mathrm{i}\omega_{m}+\gamma_{\mathbf{k}}-\mu)\delta_{\mathbf{k,k}^{\prime}} &
-U\Delta(q) & \xi_{k}\delta_{\mathbf{k,k}^{\prime}} & 0\\
-U\Delta^{\ast}(-q) & (\mathrm{i}\omega_{m}+\gamma_{\mathbf{k}}+\mu)\delta
_{\mathbf{k,k}^{\prime}} & 0 & -\xi_{k}\delta_{\mathbf{k,k}^{\prime}}\\
\xi_{k}^{\ast}\delta_{\mathbf{k,k}^{\prime}} & 0 & (\mathrm{i}\omega_{m}-\gamma
_{\mathbf{k}}-\mu)\delta_{\mathbf{k,k}^{\prime}} & -U\Delta(q)\\
0 & -\xi_{k}^{\ast}\delta_{\mathbf{k,k}^{\prime}} & -U\Delta^{\ast}(-q) &
(\mathrm{i}\omega_{m}-\gamma_{\mathbf{k}}+\mu)\delta_{\mathbf{k,k}^{\prime}}%
\end{array}
\right) .
\]
\end{widetext}

Then we consider the phase fluctuations on the SF order parameter and define
$U\Delta(q)=\Delta_{0}+\Lambda(q)$ where the mean field value $\Delta_{0\text{
}}$ is a real constant and $\Lambda(\mathbf{q})$ is complex fluctuating field.
Thus $G^{-1}$ is divided into two parts, i.e.,
\begin{equation}
G^{-1}=G_{0}^{-1}+G_{1}^{-1}%
\end{equation}
where\begin{widetext}
\[
G_{0}^{-1}=\left(
\begin{array}
[c]{cccc}%
\mathrm{i}\omega_{m}+\gamma_{\mathbf{k}}-\mu & -\Delta_{0\text{ }} & \xi_{k} &
0\\
-\Delta_{0\text{ }} & \mathrm{i}\omega_{m}+\gamma_{\mathbf{k}}+\mu & 0 &
-\xi_{k}\\
\xi_{k}^{\ast} & 0 & \mathrm{i}\omega_{m}-\gamma_{\mathbf{k}}-\mu &
-\Delta_{0\text{ }}\\
0 & -\xi_{k}^{\ast} & -\Delta_{0\text{ }} & \mathrm{i}\omega_{m}%
-\gamma_{\mathbf{k}}+\mu)
\end{array}
\right)
\]
\end{widetext}and%
\begin{equation}
G_{1}^{-1}=\left(
\begin{array}
[c]{cccc}%
0 & -\Lambda_{A}(q) & 0 & 0\\
-\Lambda_{A}^{\ast}(-q) & 0 & 0 & 0\\
0 & 0 & 0 & -\Lambda_{B}(q)\\
0 & 0 & -\Lambda_{B}^{\ast}(-q) & 0
\end{array}
\right)  .
\end{equation}
Using the expansion of the logarithm, one gets
\begin{align}
\mathbf{Tr}\ln G^{-1}  &  =\mathbf{Tr}\ln(G_{0}^{-1}+G_{1}^{-1})\\
&  =\mathbf{Tr}\ln G_{0}^{-1}-\frac{1}{2}\mathbf{Tr(}G_{0}G_{1}^{-1}G_{0}%
G_{1}^{-1})\nonumber
\end{align}

Next we investigate the Gaussian fluctuations of the paring field $\Lambda(q)$
around the saddle point $\Delta_{0}$\cite{flu,tay}. The fluctuation field is
written as%
\begin{equation}
\Lambda^{\dagger}(q)=(\Lambda_{A}^{\ast}(-q),\Lambda_{A}(q),\Lambda
_{B}(q),\Lambda_{B}^{\ast}(-q)).
\end{equation}
Then quadratic effective action becomes%
\begin{equation}
S=S_{0}+S_{1}+2\beta N\sum \limits_{q}U\Delta^{\ast}(\mathbf{q})\Delta
(\mathbf{q})+2\beta \mu N,
\end{equation}
where%
\begin{align}
S_{0}  &  =\frac{\beta}{N}\mathbf{Tr}\ln G_{0}^{-1},\\
S_{1}  &  =\frac{\beta}{2N}\sum \limits_{\mathbf{q,k}}\Lambda^{\dagger
}(\mathbf{q})Q(\mathbf{q,k})\Lambda(\mathbf{q}),
\end{align}
and the detailed form of elements in $Q(\mathbf{q,k})$ are shown in Appendix
A. Using the Matsubara summation formula, one can obtain the quantities
$Q_{i,j}$ above. Then in the static limit, i.e., $\mathrm{i}\omega
_{l}\rightarrow0$, at zero temperature, $Q_{i,j}$ can be described as follows,
for example, if
\begin{equation}
Q_{i,j}=\sum \limits_{\omega_{m}}G_{0bc}(k\mathbf{)}G_{0gh}(k-q\mathbf{)}%
\end{equation}
and then after the summation, it becomes%
\begin{align}
Q_{ij}  &  =-\frac{A_{bc}(\mathbf{k})B_{gh}(\mathbf{k-q})+B_{bc}%
(\mathbf{k})A_{gh}(\mathbf{k-q})}{e_{1,\mathbf{k-q}}+e_{1,\mathbf{k}}}\\
&  -\frac{A_{bc}(\mathbf{k})D_{gh}(\mathbf{k-q})+B_{bc}(\mathbf{k}%
)C_{gh}(\mathbf{k-q})}{e_{2,\mathbf{k-q}}+e_{1,\mathbf{k}}}\nonumber \\
&  -\frac{C_{bc}(\mathbf{k})B_{gh}(\mathbf{k-q})+D_{bc}(\mathbf{k}%
)A_{gh}(\mathbf{k-q})}{e_{1,\mathbf{k-q}}+e_{2,\mathbf{k}}}\nonumber \\
&  -\frac{C_{bc}(\mathbf{k})D_{gh}(\mathbf{k-q})+D_{bc}(\mathbf{k}%
)C_{gh}(\mathbf{k-q})}{e_{2,\mathbf{k-q}}+e_{2,\mathbf{k}}},\nonumber
\end{align}
where the parameters $A_{ij}$, $B_{ij}$, $C_{ij}$, $D_{ij}$ are all shown in
Appendix A.

In order to obtain the superfluid phase stiffness, we further separate the
fluctuation into its amplitude and phase components $\Lambda_{A/B}%
(\mathbf{q})=[\lambda_{A/B}(\mathbf{q})+\mathrm{i}\theta_{A/B}(\mathbf{q}%
)]/\sqrt{2}$ with real filed $\lambda_{A/B}(\mathbf{q})$ and $\theta
_{A/B}(\mathbf{q})$. The changes of basis can be written as%
\begin{equation}
\left(
\begin{array}
[c]{c}%
\Lambda_{A}^{\ast}(-\mathbf{q})\\
\Lambda_{A}(\mathbf{q})\\
\Lambda_{B}(\mathbf{q})\\
\Lambda_{B}^{\ast}(-\mathbf{q})
\end{array}
\right)  =\frac{1}{\sqrt{2}}\left(
\begin{array}
[c]{cccc}%
1 & \mathrm{i} & 0 & 0\\
1 & -\mathrm{i} & 0 & 0\\
0 & 0 & -\mathrm{i} & 1\\
0 & 0 & \mathrm{i} & 1
\end{array}
\right)  \left(
\begin{array}
[c]{c}%
\lambda_{A}(\mathbf{q})\\
\theta_{A}(\mathbf{q})\\
\theta_{B}(\mathbf{q})\\
\lambda_{B}(\mathbf{q})
\end{array}
\right)  .
\end{equation}
Then we have
\begin{equation}
S_{1}\mathbf{=}\frac{\beta}{2}\sum \limits_{\mathbf{q}}\lambda^{\ast}\left(
\mathbf{q}\right)  W\lambda(\mathbf{q}),
\end{equation}
where $\lambda=\left(
\begin{array}
[c]{cccc}%
\lambda_{A}(\mathbf{q}) & \theta_{A}(\mathbf{q}) & \theta_{B}(\mathbf{q}) &
\lambda_{B}(\mathbf{q})
\end{array}
\right)  ^{T}$, and the detailed forms of elements in matrix $W$ are shown in
Appendix A. Integrating over the gapped field $\lambda(\mathbf{q)}$, in the
static limit at zero temperature, we obtain%
\begin{equation}
S_{1}[\theta]=\frac{\beta}{2}\sum \limits_{\mathbf{q}}[\theta_{A}%
(\mathbf{q}),\theta_{B}(\mathbf{q})]X\left[
\begin{array}
[c]{c}%
\theta_{A}(\mathbf{q})\\
\theta_{B}(\mathbf{q})
\end{array}
\right]
\end{equation}
with%
\begin{equation}
X=\left(
\begin{array}
[c]{cc}%
W_{22}-W_{B11} & W_{23}-W_{B12}\\
W_{32}-W_{B21} & W_{33}-W_{B22}%
\end{array}
\right)  ,
\end{equation}
where the elements $W_{B11}$, $W_{B12}$, $W_{B21}$, $W_{B22}$ become
\begin{align}
W_{B11}  &  =W_{A11}W_{21}W_{12}+W_{A12}W_{21}W_{42}\\
&  +W_{A21}W_{24}W_{12}+W_{A22}W_{24}W_{42}\nonumber \\
W_{B12}  &  =W_{A11}W_{21}W_{13}+W_{A12}W_{21}W_{43}\nonumber \\
&  +W_{A21}W_{24}W_{13}+W_{A22}W_{24}W_{43}\nonumber \\
W_{B21}  &  =W_{A11}W_{31}W_{12}+W_{A12}W_{31}W_{42}\nonumber \\
&  +W_{A21}W_{34}W_{12}+W_{A22}W_{34}W_{42}\nonumber \\
W_{B22}  &  =W_{A11}W_{31}W_{13}+W_{A12}W_{31}W_{43}\nonumber \\
&  +W_{A21}W_{34}W_{13}+W_{A22}W_{34}W_{43}\nonumber
\end{align}
and the elements $W_{A11}$, $W_{A12}$, $W_{A21}$, $W_{A22}$ are given by
\begin{align}
W_{A11}  &  =\frac{W_{44}}{W_{11}W_{44}-W_{14}W_{14}}\\
W_{A12}  &  =\frac{-W_{14}}{W_{11}W_{44}-W_{14}W_{14}}\nonumber \\
W_{A21}  &  =\frac{-W_{41}}{W_{11}W_{44}-W_{14}W_{14}}\nonumber \\
W_{A22}  &  =\frac{W_{11}}{W_{11}W_{44}-W_{14}W_{14}}.\nonumber
\end{align}
At last, we arrive at the the effective action for phase fluctuations as
\begin{equation}
S[\theta]=\frac{\beta}{2}\sum \limits_{\mathbf{q}}[\theta_{A}(\mathbf{q}%
),\theta_{B}(\mathbf{q})]T\left[
\begin{array}
[c]{c}%
\theta_{A}(\mathbf{q})\\
\theta_{B}(\mathbf{q})
\end{array}
\right]
\end{equation}
where the elements of matrix $T$ read
\begin{align}
T_{11}(\mathbf{q})  &  =\frac{1}{U}+(W_{22}-W_{B11})\\
T_{12}(\mathbf{q})  &  =(W_{23}-W_{B12})\nonumber \\
T_{21}(\mathbf{q})  &  =(W_{32}-W_{B21})\nonumber \\
T_{22}(\mathbf{q})  &  =\frac{1}{U}+(W_{33}-W_{B22}).\nonumber
\end{align}
We may derive the zero temperature superfluid stiffness $\rho_{s}(0)$
numerically in the static limit by identifying\cite{flu,tay,zhao}
\begin{equation}
\sqrt{T_{11}(\mathbf{q})T_{22}(\mathbf{q})}-\sqrt{T_{12}(\mathbf{q}%
)T_{21}(\mathbf{q})}=\frac{\sqrt{3}\rho_{s}(0)}{2}\mathbf{q}^{2}%
\end{equation}
for $\mathbf{q}^{2}\rightarrow0$.

After obtaining the phase stiffness of $\mathcal{C}=\pm1$ topological SF
order, the effective Lagrangian of the phase fluctuations is obtained as
\begin{equation}
L_{p}=\frac{1}{2}\rho_{s}(0)(\mathbf{\nabla}\theta)^{2}.
\end{equation}
For example, for the case of $U=15t,$ $t^{\prime}=0.05t$, $\mu_{1}=\mu
_{2}=-h=-7t$, we have a small phase stiffness as
\[
\rho_{s}(0)\simeq0.00286t.
\]

\section{Kosterlitz-Thouless transition}

From above calculations, one can see that the induced SF pairing in layer-1
will be determined by the SF pairing in layer-2 as $\Delta_{1,\mathrm{induce}%
}=-\frac{\left(  t_{\perp}\right)  ^{2}}{U\Delta_{2}}$. If there exists a
vortex in layer-2, $\Delta_{2,i}\rightarrow \Delta_{2}\exp[\mathrm{i}%
\sum_{l\neq i}\operatorname{Im}\ln(z_{i}-z_{l})]$, there will appear induced
vortex in layer-1, $\Delta_{1,\mathrm{induce},i}\rightarrow \Delta
_{1,\mathrm{induce}}\exp[-\mathrm{i}\sum_{l\neq i}$\textrm{Im }$\ln
(z_{i}-z_{l})]$ where $z_{i}$ is the position as $z_{i}=x_{i}+iy_{i}.$ So we
can only study the dynamics of vortices in layer-1 which is defined as
\begin{equation}
\theta_{i}=\sum_{l\neq i}\operatorname{Im}\ln(z_{i}-z_{l})~ \label{phih}%
\end{equation}
where $z_{i}$ is the vortex position as $z_{i}=x_{i}+iy_{i}$.

According to the above analysis, we can get the Kosterlitz-Thouless (KT)
transition temperature by the renormalized group (RG) theory. For two vortices
in layer-1, there exists a confinement potential as
\begin{equation}
V\simeq q^{2}\ln \frac{\left \vert \mathbf{r}\right \vert }{a} \label{vdipole}%
\end{equation}
at $|\mathbf{r|>}a$, where $q^{2}=2\pi \rho_{s}(0)$, and $\mathbf{r}$ is the
distance between the vortex and anti-vortex. With the increase of temperature,
the vortex-antivortex pairs can be thermally excited, leading to a
contribution to the screening effect by reducing $V$ to $V_{\mathrm{eff}%
}=\frac{1}{\kappa}V,$ where $\kappa$ denotes the dielectric constant. In the
following, we shall treat the screening effect based on an RG treatment.

In an RG procedure, the contributions from the pairs with the sizes between
$r$ and $r+dr$ will be integrated out, starting from\textbf{\ }$r=a$. The
probability for the vortex-antivortex pairs separated by a distance $r$ is
controlled by the pair fugacity $y^{2}(r)$. In the KT theory \cite{BKT,kost1},
the initial is $y^{2}(a)=e^{-\beta E_{c}}$, (where $\beta=\frac{1}{k_{B}T}$
and $E_{c}$ is the core energy). The renormalization effect is then
represented by two renormalized quantities, $X(r)\equiv \frac{2\pi \kappa}{\beta
q^{2}}$ and $y^{2}(r)$, which satisfy the following famous recursion
relations
\begin{align}
dy/dl  &  =(2-\frac{\pi}{X})\,y,\label{rc2}\\
dX/dl  &  =4\pi^{3}y^{2}\text{,} \label{rc30}%
\end{align}
where $r=ae^{l}$. From Eqs.(\ref{rc2})-(\ref{rc30}), we find
\begin{equation}
y^{2}=y_{0}^{2}+\frac{1}{\pi^{3}}(X-X_{0})-\frac{1}{2\pi^{2}}\ln \frac{X}%
{X_{0}},
\end{equation}
where $X_{0}\equiv X(l=0)=\frac{2\pi}{\beta q^{2}}$ (with $\kappa(l=0)=1$).
The RG flow is then obtained from Eq.(\ref{rc30}) by
\begin{equation}
l=\int_{X_{0}}^{X}\frac{dX^{\prime}}{4\pi^{3}Y_{0}^{2}+4(X^{\prime}%
-X_{0})-2\pi \ln(X^{\prime}/X_{0})}. \label{l}%
\end{equation}
The pair fugacity can be determined by $y^{2}(l)=e^{-2\int_{0}^{l}(2-\frac
{\pi}{X})dl^{\prime}}$.

The RG flow diagram of Eqs.(\ref{rc2})-(\ref{rc30}) is as follows: the two
basins of attraction are separated by the initial values which flow to
$X^{\ast}\rightarrow \frac{\pi}{2}$ and $y^{\ast}\rightarrow0$ in the limit
$l\rightarrow \infty$. In terms of Eq.(\ref{l}), the separatrix of the RG flows
is given by
\begin{equation}
l=\int_{X_{0}}^{X}\frac{dX^{\prime}}{4(X^{\prime}-\frac{\pi}{2})-2\pi
\ln(2X^{\prime}/\pi)}. \label{rc4}%
\end{equation}
Based on the RG equation of (\ref{rc4}), one can determine the KT temperature
$T_{\mathrm{KT}}$. Finally we approximately have
\begin{equation}
X(l)\simeq X_{0}\simeq X(l\rightarrow \infty)=\frac{\pi}{2}%
\end{equation}
and
\begin{equation}
k_{B}T_{\mathrm{KT}}\simeq \frac{q^{2}}{4}=\frac{\pi \rho_{s}(0)}{2}.
\end{equation}

For the case of $U=15t,$ $t^{\prime}=0.05t$, $\mu_{1}=\mu_{2}=-h=-7t$, we
found a fairly low KT transition temperature as
\begin{equation}
k_{B}T_{\mathrm{KT}}=\frac{\pi \rho_{s}(0)}{2}\simeq0.0045t.
\end{equation}
Below $T_{\mathrm{KT}}\simeq0.0045t/k_{B},$ we have a TSF with long range
phase coherence. In the temperature region $T_{\mathrm{KT}}<T<T_{c}%
\simeq2.25t/k_{B},$ we have the SF pairing but no phase coherence. And in this
region, the vortex are deconfined from the bound state. At higher temperature,
$T>T_{c}\simeq2.25t/k_{B}$, the SF pairing order disappears.

\section{Conclusion}

In the end, we conclude our discussions. We propose a scenario in which a
topological phase, possessing gapless edge states and non-Abelian anyons, is
realized by proximity effect between a $\mathcal{C}=\pm1$ topological
insulator and an $s$-wave SF of ultracold fermionic atoms in a bilayer optical
lattice with an effective gauge field and a layer-dependent Zeeman field
generated by laser-field. At the beginning, we give an effective design of the
bilayer Haldane model. Then we put two-component (two pseudo-spins)
interacting fermions on this bilayer optical lattice with fixed particle
concentration. For layer-1, the Haldane model of two-component fermions at
$1/4$ filling under a strong Zeeman field becomes a $\mathcal{C}=\pm1$
topological insulator. While for layer-2, there is no Zeeman fields, we get an
s-wave SF state by tuning the interaction between fermion via Feshbach
resonance technique. Thus due to the proximity effect, we get an effective
$\mathcal{C}=\pm1$ TSF state. We also study its topological properties and
then show the gapless Majorana edge modes and the non-Abelian statistics of
the $\pi$-flux. This $\mathcal{C}=\pm1$ TSF therefore may be a possible
candidate for topological quantum computation. Finally we calculate the phase
stiffness by using the RPA approach and then derive the temperature of the KT
transition for the system.

\begin{acknowledgments}
The authors thank W. Yi and Z. W. Zhou for their helpful discussion. This work is supported by NFSC Grant No. 11174035, National Basic Research
Program of China (973 Program) under the grant No. 2011CB921803, 2012CB921704.
\end{acknowledgments}

\appendix

\section{Parameters in Green functions}

In the appendix, we first give the elements in $Q(\mathbf{q,k})$:
\begin{align}
Q_{11}  &  =\frac{1}{\beta}\sum \limits_{\omega_{m}}G_{011}(k\mathbf{)}%
G_{022}(k-q\mathbf{)}\\
Q_{12}  &  =\frac{1}{\beta}\sum \limits_{\omega_{m}}G_{012}(k\mathbf{)}%
G_{012}(k-q\mathbf{)}\nonumber \\
Q_{13}  &  =\frac{1}{\beta}\sum \limits_{\omega_{m}}G_{014}(k\mathbf{)}%
G_{032}(k-q\mathbf{)}\nonumber \\
Q_{14}  &  =\frac{1}{\beta}\sum \limits_{\omega_{m}}G_{013}(k\mathbf{)}%
G_{042}(k-q\mathbf{)}\nonumber
\end{align}%
\begin{align}
Q_{21}  &  =\frac{1}{\beta}\sum \limits_{\omega_{m}}G_{021}(k\mathbf{)}%
G_{021}(k-q\mathbf{)}\\
Q_{22}  &  =\frac{1}{\beta}\sum \limits_{\omega_{m}}G_{022}(k\mathbf{)}%
G_{011}(k-q\mathbf{)}\nonumber \\
Q_{23}  &  =\frac{1}{\beta}\sum \limits_{\omega_{m}}G_{024}(k\mathbf{)}%
G_{031}(k-q\mathbf{)}\nonumber \\
Q_{24}  &  =\frac{1}{\beta}\sum \limits_{\omega_{m}}G_{023}(k\mathbf{)}%
G_{041}(k-q\mathbf{)}\nonumber
\end{align}%
\begin{align}
Q_{31}  &  =\frac{1}{\beta}\sum \limits_{\omega_{m}}G_{041}(k\mathbf{)}%
G_{023}(k-q\mathbf{)}\\
Q_{32}  &  =\frac{1}{\beta}\sum \limits_{\omega_{m}}G_{042}(k\mathbf{)}%
G_{013}(k-q\mathbf{)}\nonumber \\
Q_{33}  &  =\frac{1}{\beta}\sum \limits_{\omega_{m}}G_{044}(k\mathbf{)}%
G_{033}(k-q\mathbf{)}\nonumber \\
Q_{34}  &  =\frac{1}{\beta}\sum \limits_{\omega_{m}}G_{043}(k\mathbf{)}%
G_{043}(k-q\mathbf{)}\nonumber
\end{align}%
\begin{align}
Q_{41}  &  =\frac{1}{\beta}\sum \limits_{\omega_{m}}G_{031}(k\mathbf{)}%
G_{024}(k-q\mathbf{)}\\
Q_{42}  &  =\frac{1}{\beta}\sum \limits_{\omega_{m}}G_{032}(k\mathbf{)}%
G_{014}(k-q\mathbf{)}\nonumber \\
Q_{43}  &  =\frac{1}{\beta}\sum \limits_{\omega_{m}}G_{034}(k\mathbf{)}%
G_{034}(k-q\mathbf{)}\nonumber \\
Q_{44}  &  =\frac{1}{\beta}\sum \limits_{\omega_{m}}G_{033}(k\mathbf{)}%
G_{044}(k-q\mathbf{).}\nonumber
\end{align}
Here we have
\begin{equation}
G_{0ij}(\mathbf{k)=}\frac{A_{ij}}{\mathrm{i}\omega_{m}-e_{_{1\mathbf{,k}}}%
}+\frac{B_{ij}}{\mathrm{i}\omega_{m}+e_{_{1\mathbf{,k}}}}+\frac{C_{ij}%
}{\mathrm{i}\omega_{m}-e_{_{2\mathbf{,k}}}}+\frac{D_{ij}}{\mathrm{i}\omega
_{m}+e_{_{2\mathbf{,k}}}},
\end{equation}
where the parameters $A_{ij}$, $B_{ij}$, $C_{ij}$, $D_{ij}$ in $G_{0ij}%
(\mathbf{k)}$ are given by\
\begin{align}
A_{11}(\mathbf{k})  &  =\frac{1+p_{11}(\mathbf{k})}{4}-\frac{\gamma
_{\mathbf{k}}-\mu+q_{11}(\mathbf{k})}{4e_{1}}\\
B_{11}(\mathbf{k})  &  =\frac{1+p_{11}(\mathbf{k})}{4}+\frac{\gamma
_{\mathbf{k}}-\mu+q_{11}(\mathbf{k})}{4e_{1}}\nonumber \\
C_{11}(\mathbf{k})  &  =\frac{1-p_{11}(\mathbf{k})}{4}-\frac{\gamma
_{\mathbf{k}}-\mu-q_{11}(\mathbf{k})}{4e_{1}}\nonumber \\
D_{11}(\mathbf{k})  &  =\frac{1-p_{11}(\mathbf{k})}{4}+\frac{\gamma
_{\mathbf{k}}-\mu+q_{11}(\mathbf{k})}{4e_{1}}\nonumber
\end{align}
with%
\begin{align}
p_{11}(\mathbf{k})  &  =\frac{-\mu \gamma_{\mathbf{k}}}{\sqrt{\gamma
_{\mathbf{k}}^{2}[\mu^{2}+\Delta_{0\text{ }}^{2}]+|\xi_{\mathbf{k}}|^{2}%
\mu^{2}}}\\
q_{11}(\mathbf{k})  &  =\frac{\gamma_{\mathbf{k}}(\mu^{2}+\Delta_{0\text{ }%
}^{2})-\mu(\gamma_{\mathbf{k}}^{2}+|\xi_{\mathbf{k}}|^{2})}{\sqrt
{\gamma_{\mathbf{k}}^{2}[\mu^{2}+\Delta_{0\text{ }}^{2}]+|\xi_{\mathbf{k}%
}|^{2}\mu^{2}}}\nonumber
\end{align}
and%
\begin{align}
A_{12}(\mathbf{k})  &  =\frac{p_{12}(\mathbf{k})}{4}+\frac{\Delta_{0\text{ }%
}-q_{12}(\mathbf{k})}{4e_{1}}\\
B_{12}(\mathbf{k})  &  =\frac{p_{12}(\mathbf{k})}{4}-\frac{\Delta_{0\text{ }%
}-q_{12}(\mathbf{k})}{4e_{1}}\nonumber \\
C_{12}(\mathbf{k})  &  =\frac{-p_{12}(\mathbf{k})}{4}+\frac{\Delta_{0}%
+q_{12}(\mathbf{k})}{4e_{2}}\nonumber \\
D_{12}(\mathbf{k})  &  =\frac{-p_{12}(\mathbf{k})}{4}-\frac{\Delta_{0}%
+q_{12}(\mathbf{k})}{4e_{2}}\nonumber
\end{align}
with%
\begin{align}
p_{12}(\mathbf{k})  &  =\frac{-\gamma_{\mathbf{k}}\Delta_{0\text{ }}}%
{\sqrt{\gamma_{\mathbf{k}}^{2}[\mu^{2}+\Delta_{0\text{ }}^{2}]+|\xi
_{\mathbf{k}}|^{2}\mu^{2}}}\\
q_{12}(\mathbf{k})  &  =\frac{-\Delta_{0\text{ }}\gamma_{k}^{2}}{\sqrt
{\gamma_{\mathbf{k}}^{2}[\mu^{2}+\Delta_{0\text{ }}^{2}]+|\xi_{\mathbf{k}%
}|^{2}\mu^{2}}}\nonumber
\end{align}
and%
\begin{align}
A_{13}(\mathbf{k})  &  =\frac{p_{13}(\mathbf{k})}{4}-\frac{\xi_{\mathbf{k}%
}+q_{13}(\mathbf{k})}{4e_{1}}\\
B_{13}(\mathbf{k})  &  =\frac{p_{13}(\mathbf{k})}{4}+\frac{\xi_{\mathbf{k}%
}+q_{13}(\mathbf{k})}{4e_{1}}\nonumber \\
C_{13}(\mathbf{k})  &  =\frac{-p_{13}(\mathbf{k})}{4}-\frac{\xi_{\mathbf{k}%
}-q_{13}(\mathbf{k})}{4e_{2}}\nonumber \\
D_{13}(\mathbf{k})  &  =\frac{-p_{13}(\mathbf{k})}{4}+\frac{\xi_{\mathbf{k}%
}-q_{13}(\mathbf{k})}{4e_{2}}\nonumber
\end{align}
with%
\begin{align}
p_{13}(\mathbf{k})  &  =\frac{-\mu \xi_{\mathbf{k}}}{\sqrt{\gamma_{\mathbf{k}%
}^{2}[\mu^{2}+\Delta_{0\text{ }}^{2}]+|\xi_{\mathbf{k}}|^{2}\mu^{2}}}\\
q_{13}(\mathbf{k})  &  =\frac{\xi_{\mathbf{k}}\mu^{2}}{\sqrt{\gamma
_{\mathbf{k}}^{2}[\mu^{2}+\Delta_{0\text{ }}^{2}]+|\xi_{\mathbf{k}}|^{2}%
\mu^{2}}}\nonumber
\end{align}
and%
\begin{align}
A_{14}(\mathbf{k})  &  =\frac{-q_{14}(\mathbf{k})}{4e_{1}}\\
B_{14}(\mathbf{k})  &  =\frac{q_{14}(\mathbf{k})}{4e_{1}}\nonumber \\
C_{14}(\mathbf{k})  &  =\frac{q_{14}(\mathbf{k})}{4e_{2}}\nonumber \\
D_{14}(\mathbf{k})  &  =\frac{-q_{14}(\mathbf{k})}{4e_{2}}\nonumber
\end{align}
with%
\begin{equation}
q_{14}(\mathbf{k})=\frac{\xi_{\mathbf{k}}\Delta_{0\text{ }}(\gamma
_{\mathbf{k}}+\mu)}{\sqrt{\gamma_{\mathbf{k}}^{2}[\mu^{2}+\Delta_{0\text{ }%
}^{2}]+|\xi_{\mathbf{k}}|^{2}\mu^{2}}}%
\end{equation}
and
\begin{align}
A_{21}(\mathbf{k})  &  =A_{12}^{\ast}(\mathbf{k})\\
B_{21}(\mathbf{k})  &  =B_{12}^{\ast}(\mathbf{k})\nonumber \\
C_{21}(\mathbf{k})  &  =C_{12}^{\ast}(\mathbf{k})\nonumber \\
D_{21}(\mathbf{k})  &  =D_{12}^{\ast}(\mathbf{k})\nonumber
\end{align}
and
\begin{align}
A_{22}(\mathbf{k})  &  =\frac{1+p_{22}(\mathbf{k})}{4}-\frac{\gamma
_{\mathbf{k}}+\mu+q_{22}(\mathbf{k})}{4e_{1}}\\
B_{22}(\mathbf{k})  &  =\frac{1+p_{22}(\mathbf{k})}{4}+\frac{\gamma
_{\mathbf{k}}+\mu+q_{22}(\mathbf{k})}{4e_{1}}\nonumber \\
C_{22}(\mathbf{k})  &  =\frac{1-p_{22}(\mathbf{k})}{4}-\frac{\gamma
_{\mathbf{k}}+\mu-q_{22}(\mathbf{k})}{4e_{2}}\nonumber \\
D_{22}(\mathbf{k})  &  =\frac{1-p_{22}(\mathbf{k})}{4}+\frac{\gamma
_{\mathbf{k}}+\mu-q_{22}(\mathbf{k})}{4e_{2}}\nonumber
\end{align}
with%
\begin{align}
p_{22}(\mathbf{k})  &  =\frac{\mu \gamma_{\mathbf{k}}}{\sqrt{\gamma
_{\mathbf{k}}^{2}[\mu^{2}+\Delta_{0\text{ }}^{2}]+|\xi_{\mathbf{k}}|^{2}%
\mu^{2}}}\\
q_{22}(\mathbf{k})  &  =\frac{\gamma_{\mathbf{k}}(\mu^{2}+\Delta_{0\text{ }%
}^{2})+\mu(\gamma_{k}^{2}+|\xi_{\mathbf{k}}|^{2})}{\sqrt{\gamma_{\mathbf{k}%
}^{2}[\mu^{2}+\Delta_{0\text{ }}^{2}]+|\xi_{\mathbf{k}}|^{2}\mu^{2}}}\nonumber
\end{align}
and%
\begin{align}
A_{23}(\mathbf{k})  &  =\frac{-q_{23}(\mathbf{k})}{4e_{1}}\\
B_{23}(\mathbf{k})  &  =\frac{q_{23}(\mathbf{k})}{4e_{1}}\nonumber \\
C_{23}(\mathbf{k})  &  =\frac{q_{23}(\mathbf{k})}{4e_{2}}\nonumber \\
D_{23}(\mathbf{k})  &  =\frac{-q_{23}(\mathbf{k})}{4e_{2}}\nonumber
\end{align}
with%
\begin{equation}
q_{23}(\mathbf{k})=\frac{-\xi \Delta_{0\text{ }}(\gamma_{\mathbf{k}}-\mu
)}{\sqrt{\gamma_{\mathbf{k}}^{2}[\mu^{2}+\Delta_{0\text{ }}^{2}]+|\xi
_{\mathbf{k}}|^{2}\mu^{2}}}%
\end{equation}
and
\begin{align}
A_{24}(\mathbf{k})  &  =\frac{p_{24}(\mathbf{k})}{4}+\frac{\xi_{\mathbf{k}%
}-q_{24}(\mathbf{k})}{4e_{1}}\\
B_{24}(\mathbf{k})  &  =\frac{p_{24}(\mathbf{k})}{4}-\frac{\xi_{\mathbf{k}%
}-q_{24}(\mathbf{k})}{4e_{1}}\nonumber \\
C_{24}(\mathbf{k})  &  =\frac{-p_{24}(\mathbf{k})}{4}+\frac{\xi_{\mathbf{k}%
}+q_{24}(\mathbf{k})}{4e_{2}}\nonumber \\
D_{24}(\mathbf{k})  &  =\frac{-p_{24}(\mathbf{k})}{4}-\frac{\xi_{\mathbf{k}%
}+q_{24}(\mathbf{k})}{4e_{2}}\nonumber
\end{align}
with%
\begin{align}
p_{24}(\mathbf{k})  &  =\frac{-\mu \xi_{\mathbf{k}}}{\sqrt{\gamma_{\mathbf{k}%
}^{2}[\mu^{2}+\Delta_{0\text{ }}^{2}]+|\xi_{\mathbf{k}}|^{2}\mu^{2}}}\\
q_{24}(\mathbf{k})  &  =\frac{-\xi_{\mathbf{k}}\mu^{2}}{\sqrt{\gamma
_{\mathbf{k}}^{2}[\mu^{2}+\Delta_{0\text{ }}^{2}]+|\xi_{\mathbf{k}}|^{2}%
\mu^{2}}}\nonumber
\end{align}
and
\begin{align}
A_{31}(\mathbf{k})  &  =A_{13}^{\ast}(\mathbf{k})\\
B_{31}(\mathbf{k})  &  =B_{13}^{\ast}(\mathbf{k})\nonumber \\
C_{31}(\mathbf{k})  &  =C_{13}^{\ast}(\mathbf{k})\nonumber \\
D_{31}(\mathbf{k})  &  =D_{13}^{\ast}(\mathbf{k})\nonumber
\end{align}
and
\begin{align}
A_{32}(\mathbf{k})  &  =A_{23}^{\ast}(\mathbf{k})\\
B_{32}(\mathbf{k})  &  =B_{23}^{\ast}(\mathbf{k})\nonumber \\
C_{32}(\mathbf{k})  &  =C_{23}^{\ast}(\mathbf{k})\nonumber \\
D_{32}(\mathbf{k})  &  =D_{23}^{\ast}(\mathbf{k})\nonumber
\end{align}
and%
\begin{align}
A_{33}(\mathbf{k})  &  =\frac{1+p_{33}(\mathbf{k})}{4}+\frac{\gamma
_{\mathbf{k}}+\mu-q_{33}(\mathbf{k})}{4e_{1}}\\
B_{33}(\mathbf{k})  &  =\frac{1+p_{33}(\mathbf{k})}{4}-\frac{\gamma
_{\mathbf{k}}+\mu-q_{33}(\mathbf{k})}{4e_{1}}\nonumber \\
C_{33}(\mathbf{k})  &  =\frac{1-p_{33}(\mathbf{k})}{4}+\frac{\gamma
_{\mathbf{k}}+\mu+q_{33}(\mathbf{k})}{4e_{2}}\nonumber \\
D_{33}(\mathbf{k})  &  =\frac{1-p_{33}(\mathbf{k})}{4}-\frac{\gamma
_{\mathbf{k}}+\mu+q_{33}(\mathbf{k})}{4e_{2}}\nonumber
\end{align}
with%
\begin{align}
p_{33}(\mathbf{k})  &  =\frac{\mu \gamma_{\mathbf{k}}}{\sqrt{\gamma
_{\mathbf{k}}^{2}[\mu^{2}+\Delta_{0\text{ }}^{2}]+|\xi_{\mathbf{k}}|^{2}%
\mu^{2}}}\\
q_{33}(\mathbf{k})  &  =\frac{-\gamma_{\mathbf{k}}(\mu^{2}+\Delta_{0\text{ }%
}^{2})-\mu(\gamma_{\mathbf{k}}^{2}+|\xi_{\mathbf{k}}|^{2})}{\sqrt
{\gamma_{\mathbf{k}}^{2}[\mu^{2}+\Delta_{0\text{ }}^{2}]+|\xi_{\mathbf{k}%
}|^{2}\mu^{2}}}\nonumber
\end{align}
and
\begin{align}
A_{34}(\mathbf{k})  &  =\frac{p_{34}(\mathbf{k})}{4}+\frac{\Delta_{0\text{ }%
}-q_{34}(\mathbf{k})}{4e_{1}}\\
B_{34}(\mathbf{k})  &  =\frac{p_{34}(\mathbf{k})}{4}-\frac{\Delta_{0\text{ }%
}-q_{34}(\mathbf{k})}{4e_{1}}\nonumber \\
C_{34}(\mathbf{k})  &  =\frac{-p_{34}(\mathbf{k})}{4}+\frac{\Delta_{0\text{ }%
}+q_{34}(\mathbf{k})}{4e_{2}}\nonumber \\
D_{34}(\mathbf{k})  &  =\frac{-p_{34}(\mathbf{k})}{4}-\frac{\Delta_{0\text{ }%
}+q_{34}(\mathbf{k})}{4e_{2}}\nonumber
\end{align}
with%
\begin{align}
p_{34}(\mathbf{k})  &  =\frac{\Delta_{0\text{ }}\gamma_{\mathbf{k}}}%
{\sqrt{\gamma_{\mathbf{k}}^{2}[\mu^{2}+\Delta_{0\text{ }}^{2}]+|\xi
_{\mathbf{k}}|^{2}\mu^{2}}}\\
q_{34}(\mathbf{k})  &  =\frac{-\Delta_{0\text{ }}\gamma_{\mathbf{k}}^{2}%
}{\sqrt{\gamma_{\mathbf{k}}^{2}[\mu^{2}+\Delta_{0\text{ }}^{2}]+|\xi
_{\mathbf{k}}|^{2}\mu^{2}}}\nonumber
\end{align}
and%
\begin{align}
A_{41}(\mathbf{k})  &  =A_{14}^{\ast}(\mathbf{k})\\
B_{41}(\mathbf{k})  &  =B_{14}^{\ast}(\mathbf{k})\nonumber \\
C_{41}(\mathbf{k})  &  =C_{14}^{\ast}(\mathbf{k})\nonumber \\
D_{41}(\mathbf{k})  &  =D_{14}^{\ast}(\mathbf{k})\nonumber
\end{align}
and%
\begin{align}
A_{42}(\mathbf{k})  &  =A_{24}^{\ast}(\mathbf{k})\\
B_{42}(\mathbf{k})  &  =B_{24}^{\ast}(\mathbf{k})\nonumber \\
C_{42}(\mathbf{k})  &  =C_{24}^{\ast}(\mathbf{k})\nonumber \\
D_{42}(\mathbf{k})  &  =D_{24}^{\ast}(\mathbf{k})\nonumber
\end{align}
and%
\begin{align}
A_{43}(\mathbf{k})  &  =A_{34}^{\ast}(\mathbf{k})\\
B_{43}(\mathbf{k})  &  =B_{34}^{\ast}(\mathbf{k})\nonumber \\
C_{43}(\mathbf{k})  &  =C_{34}^{\ast}(\mathbf{k})\nonumber \\
D_{43}(\mathbf{k})  &  =D_{34}^{\ast}(\mathbf{k})\nonumber
\end{align}
and
\begin{align}
A_{44}(\mathbf{k})  &  =\frac{1+p_{44}(\mathbf{k})}{4}-\frac{-\gamma
_{\mathbf{k}}+\mu+q_{44}(\mathbf{k})}{4e_{1}}\\
B_{44}(\mathbf{k})  &  =\frac{1+p_{44}(\mathbf{k})}{4}+\frac{-\gamma
_{\mathbf{k}}+\mu+q_{44}(\mathbf{k})}{4e_{1}}\nonumber \\
C_{44}(\mathbf{k})  &  =\frac{1-p_{44}(\mathbf{k})}{4}-\frac{-\gamma
_{\mathbf{k}}+\mu-q_{44}(\mathbf{k})}{4e_{2}}\nonumber \\
D_{44}(\mathbf{k})  &  =\frac{1-p_{44}(\mathbf{k})}{4}+\frac{-\gamma
_{\mathbf{k}}+\mu-q_{44}(\mathbf{k})}{4e_{2}}\nonumber
\end{align}
with%
\begin{align}
p_{44}(\mathbf{k})  &  =\frac{-\mu \gamma_{\mathbf{k}}}{\sqrt{\gamma
_{\mathbf{k}}^{2}[\mu^{2}+\Delta_{0\text{ }}^{2}]+|\xi_{\mathbf{k}}|^{2}%
\mu^{2}}}\\
q_{44}(\mathbf{k})  &  =\frac{-\gamma_{\mathbf{k}}(\mu^{2}+\Delta_{0\text{ }%
}^{2})+\mu(\gamma_{\mathbf{k}}^{2}+|\xi_{\mathbf{k}}|^{2})}{\sqrt
{\gamma_{\mathbf{k}}^{2}[\mu^{2}+\Delta_{0\text{ }}^{2}]+|\xi_{\mathbf{k}%
}|^{2}\mu^{2}}}.\nonumber
\end{align}

The detailed forms of elements in matrix $M$ are given as follows:%
\begin{align}
W_{11}  &  =\frac{1}{2N}\sum \limits_{\mathbf{k}}(Q_{11}+Q_{21}+Q_{12}%
+Q_{22})\\
W_{12}  &  =\frac{1}{2N}\sum \limits_{\mathbf{k}}i(Q_{11}+Q_{21})-i(Q_{12}%
+Q_{22})\nonumber \\
W_{13}  &  =\frac{1}{2N}\sum \limits_{\mathbf{k}}-i(Q_{13}+Q_{23}%
)+i(Q_{14}+Q_{24})\nonumber \\
W_{14}  &  =\frac{1}{2N}\sum \limits_{\mathbf{k}}(Q_{13}+Q_{23}+Q_{14}%
+Q_{24})\nonumber
\end{align}%
\begin{align}
W_{21}  &  =\frac{1}{2N}\sum \limits_{\mathbf{k}}-i(Q_{11}-Q_{21}%
)-i(Q_{12}-Q_{22})\\
W_{22}  &  =\frac{1}{2N}\sum \limits_{\mathbf{k}}(Q_{11}-Q_{21}-Q_{12}%
+Q_{22})\nonumber \\
W_{23}  &  =\frac{1}{2N}\sum \limits_{\mathbf{k}}(-Q_{13}+Q_{23}+Q_{14}%
-Q_{24})\nonumber \\
W_{24}  &  =\frac{1}{2N}\sum \limits_{\mathbf{k}}-i(Q_{13}-Q_{23}%
)-i(Q_{14}-Q_{24})\nonumber
\end{align}%
\begin{align}
W_{31}  &  =\frac{1}{2N}\sum \limits_{\mathbf{k}}i(Q_{31}-Q_{41})+i(Q_{32}%
-Q_{42})\\
W_{32}  &  =\frac{1}{2N}\sum \limits_{\mathbf{k}}(-Q_{31}+Q_{32}+Q_{41}%
-Q_{42})\nonumber \\
W_{33}  &  =\frac{1}{2N}\sum \limits_{\mathbf{k}}(Q_{33}-Q_{43}-Q_{34}%
-Q_{44})\nonumber \\
W_{34}  &  =\frac{1}{2N}\sum \limits_{\mathbf{k}}i(Q_{33}-Q_{43})+i(Q_{34}%
-Q_{44})\nonumber
\end{align}%
\begin{align}
W_{41}  &  =\frac{1}{2N}\sum \limits_{\mathbf{k}}(Q_{31}+Q_{41}+Q_{32}%
+Q_{42})\\
W_{42}  &  =\frac{1}{2N}\sum \limits_{\mathbf{k}}i(Q_{31}+Q_{41})-i(Q_{32}%
+Q_{42})\nonumber \\
W_{43}  &  =\frac{1}{2N}\sum \limits_{\mathbf{k}}-i(Q_{33}+Q_{43}%
)+i(Q_{34}+Q_{44})\nonumber \\
W_{44}  &  =\frac{1}{2N}\sum \limits_{\mathbf{k}}(Q_{33}+Q_{43}+Q_{34}%
+Q_{44}).\nonumber
\end{align}

\end{document}